\documentstyle[12pt]{article}
\renewcommand{\theequation}{\thesection.\arabic{equation}}
\makeatletter
\def\eqnarray{
\stepcounter{equation}\let\@currentlabel=\theequation
\global\@eqnswtrue
\global\@eqcnt\z@\tabskip\@centering\let\\=\@eqncr
$$\halign to \displaywidth\bgroup\@eqnsel\hskip\@centering
  $\displaystyle\tabskip\z@{##}$&\global\@eqcnt\@ne 
  \hfil$\displaystyle{{}##{}}$\hfil
  &\global\@eqcnt\tw@$\displaystyle\tabskip\z@{##}$\hfil
  \tabskip\@centering&\llap{##}\tabskip\z@\cr}
\@addtoreset{equation}{section}
\makeatother
\begin{document}
%
%
\begin{center}
\large{The Extended Nambu-Jona-Lasinio Model and}\\
\large{Hidden Local Symmetry of Low Energy QCD}
\end{center}
\vspace{6mm}
\begin{center}
M. Wakamatsu
\end{center}
\vspace{1mm}
\begin{center}
Department of Physics, Faculty of Science, 
\end{center}
\begin{center}
Osaka University, 
\end{center}
\begin{center}
Toyonaka, Osaka 560, JAPAN
\end{center}
\vspace{4mm}
\begin{flushleft}
PACS numbers \ : \ 11.30.-j, \,\,12.39.Fe, \,\,12.40.Vv
\end{flushleft}
\vspace{4mm}
\paragraph{Abstract :}

Using the standard auxiliary field method, we derive from the extended
Nambu-Jona-Lasinio model an effective meson action containing
vector and axial-vector mesons in addition to Goldstone bosons.
The vector and axial-vector mesons in this effective action transform as
gauge fields of hidden local symmetry $G_{local} = {[U(n)_L \times 
U(n)_R]}_{local}$. Here, the realization of enlarged hidden local symmetry
is accomplished via the introduction of two kinds of ``compensating''
fields. For obtaining the intrinsic-parity violating part of the action, we
generalize the standard gauged Wess-Zumino-Witten action such that it also
contains two kinds of ``compensators'' in addition to the usual Goldstone
bosons as well as the vector and axial-vector mesons. This generalized
gauged Wess-Zumino-Witten action turns out to have $G_{globa} \times
G_{local}$ symmetry, where $G_{global}$ being the usual $U(n)_L \times
U(n)_R$ global chiral symmetry while $G_{local}$ being the
$U(n)_L \times U(n)_R$ hidden local symmetry.
This means that $G_{local}$ has no gauge anomaly and its associated vector
and axial-vector mesons can be regarded as gauge bosons of $G_{local}$.
The introduction
of the coupling with the external electroweak fields requires us to
gauge some appropriate subgroup of $G_{global}$. To perform it in
consistent with the anomaly structure of QCD is a nontrivial problem.
We explain how this can be done, following the recent suggestion by
several authors.

\vspace{8mm}
\section{Introduction}

\ \ \ \ \ \ It is a widely-accepted belief that the applicable region
of the nonlinear sigma model as a low energy effective theory of QCD
can be extended to higher energies by incorporating other mesons
(especially the vector and axial-vector mesons) than
the Nambu-Goldstone bosons as explicit dynamical degrees of freedom.
There are several ways to introduce spin-1 mesons into the basic chiral
lagrangian [1-6]. Widely known examples includes the so-called massive
Yang-Mills scheme as well as the scheme based on the hidden local
symmetry initiated by Bando et al. [6-9].
Also known for a long time is a general theoretical framework
based on the nonlinear realization of chiral symmetry initiated by
Weinberg [10] and further developed by
Callan, Coleman, Wess and Zumino [11]. Another approach, in which spin-1
mesons are represented as antisymmetric tensor fields, have also been
proposed recently [12].
Although it is a general belief now that all of these
approaches are in principle equivalent (see, for instance, the recent
review by Birse [4]), the scheme based on the hidden gauge symmetry
has attracted special attention because of its several appealing
features [5,6]. Bando et al's original construction of the model
is based on the observation that a nonlinear sigma model based on the
manifold $G \,/ \,H \,= \,U(3)_L \times U(3)_R \,/ \,U(n)_V$ is gauge
equivalent to a ``linear'' model with $G_{global} \times H_{local}$
symmetry, where $H_{local}$ is the hidden local symmetry whose
corresponding gauge fields are composite gauge bosons [7,13,14].
Later, Bando et al. enlarged the hidden local symmetry further
into $G_{local} = U(3)_L \times U(3)_R$, which enables them to construct
an effective lagrangian containing not only the vector mesons but also
the axial-vector mesons [8,9].
A basic assumption in their construction is
that the kinetic terms of such composite gauge bosons are generated
through some quantum effects, as it actually happens for the $CP_{N-1}$
model [15]. Up to now, the
validity of this scenario has been explicitly confirmed only within the
extended Nambu-Jona-Lasinio (NJL) model as a tractable substitute of
QCD lagrangian in the low energy domain [16-24].
In fact, the generation mechanism of the kinetic terms of
spin-1 bosons has long been known in the auxiliary field treatment of
the extended NJL lagrangian [25-27].
Some years ago, we have shown that an approximate
bosonisation of the extended NJL lagrangian by using the auxiliary field
method leads to a gauge fixed form of Bando et al's lagrangian with
the enlarged hidden local symmetry $U(n)_L \times U(n)_R$ [21].
A natural question is whether one can also obtain a corresponding
lagrangian with explicit hidden gauge symmetry. It turns out that it
is in fact possible if one introduces two kinds of compensating fields
(or ``compensators'') ,which plays the role of the gauge parameters to be
absorbed into the masses of the vector and axial-vector mesons in the
unitary gauge.

In this paper, we shall explicitly derive an effective meson lagrangian
with the enlarged hidden local symmetry belonging to
$U(n)_L \times U(n)_R$, by starting from the extended NJL model.
Due to the presence of the $\gamma^\mu \,\gamma_5$ coupling between the
quark fields and the auxiliary axial-vector fields, we must pay special
care to symmetries possessed by the original lagrangian, which may not be
maintained in the resultant effective meson action due to the
chiral anomaly. The original NJL lagrangian has global chiral
symmetry. Since this is also a fundamental symmetry of strong interactions,
we want to keep it unbroken after quark loop integral.
On the other hand, the hidden local symmetry
$G_{local} = {[U(n)_L \times U(n)_R]}_{local}$ is put into the formalism
by introducing two kinds of compensating fields [6,9].
We nevertheless want to maintain this symmetry, since we can then
obtain an effective gauge theory of vector and axial-vector mesons.
The question is now whether it is possible to maintain both of these
symmetries simultaneously.
(That this is not a trivial question may be deduced
from Bando et al's remark in [6] that in the case
$G_{local} = {[U(n)_L \times U(n)_R]}_{local}$, the anomaly associated with
$G_{local}$ should be canceled by an {\it extra} Wess-Zumino term, since
QCD possesses the $G_{global}$ anomaly but not the $G_{local}$ anomaly
at all.) As we shall see, if it were not for the couplings with the
external electroweak gauge fields, an effective action with the desired
symmetries can readily be obtained. However, once these couplings are
introduced, a nontrivial problem arises, which
has in fact caused much confusion in the past [28-31].
It seemed that within the
framework of the extended NJL model there is no way to satisfy both
conditions, i.e. the global chiral symmetry of hadronic processes and
the electromagnetic gauge invariance [22,23]. Recently, a solution to this
problem has been proposed by Bijnens and Prades [32]. (See also [33].)
According to them, there is some uncertainty in the way the four
quark vertex in the extended NJL model is treated. To be more explicit,
they pointed out that the standardly assumed choice of the path integral
measure corresponding to the hadronic vector and axial-vector fields are
not necessarily justified. Their observation opens up a possibility
to subtract local counter terms, which depend on both of the hadronic
vector and axial-vector fields and of the external electroweak gauge
fields, to obtain an anomalous action with the desired symmetry.
Making use of this observation, we can in fact obtain an anomalous action,
which respects the global chiral symmetry at the strong interaction level
as well as the electromagnetic gauge invariance, while keeping the
full hidden local symmetry. To show that it is in fact possible is the
main purpose of the present paper. We believe that this explicit
construction will help us to deepen our understanding about
the meaning of the hidden local symmetry in low energy effective
theories of QCD.

The plan of the paper is as follows. In sect.2, we treat the case in which
the electroweak couplings are switched off. The realistic case with the
electroweak couplings will be discussed in sect 3. Sect.4 summarizes
main results of the present study.
Possible advantages in working in a theory with
extra gauge degrees of freedom will also be discussed there.

\vspace{4mm}
\setcounter{equation}{0}
\section{Extendend NJL model and its effective meson action}

\subsection{Definition of effective meson action}

\ \ \ \ \ \ Here we start with the following extended NJL model with
its chirally invariant four-fermion couplings [16-24] :
\begin{eqnarray}
   {\cal L}_{NJL} \ &=& \ \bar{q} \,i \,\gamma^\mu \,\partial_\mu \,q
   \ \ + \ \ 2 \,\,G_S \,\,\sum_{a=1}^{n^2 - 1} \,\,\{ \,
   {( \bar{q} \,\,T^a \,q)}^2 \ + \ 
   {( \bar{q} \,\,i \,\gamma_5 \,T^a \,q)}^2
   \, \} \nonumber \\
   &\,& \hspace{22mm} - \ \ \,2 \,\,G_V \,\,\sum_{a=1}^{n^2 - 1} \,\,\{ \,
   {( \bar{q} \,\,\gamma^\mu \,T^a \,q)}^2 \ + \ 
   {( \bar{q} \,\,\gamma^\mu \,\gamma_5 \,T^a \,q)}^2
   \, \} \,\,.
\end{eqnarray}
Here $q$ are the quark fields, $n$ is the number of the flavor degrees
of freedom, and $T^a$ are generators of the flavor $U(n)$ group normalized
as $\mbox{tr} \,(T^a T^b) = \frac{1}{2} \,\delta_{ab}$. (The color indices
of quarks are not shown explicitly.)
Throughout the present study, we shall
neglect the bare quark masses, for simplicity. In this chiral limit,
the above lagrangian has exact chiral symmetry
${[U(n)_L \times U(n)_R]}_{global}$. (We also neglect the so-called
$U_A (1)$ problem for reasons of simplicity [34].)

Introducing the color singlet collective (auxiliary) meson fields in the
standard way, the lagrangian (2.1) can be rewritten as follows.
First define ${\cal L}$ by
\begin{eqnarray}
   {\cal L} \ \ = \ \ {\cal L}_{NJL} \ \ + \ \ {\cal L}_{auxiliary} \,\, ,
\end{eqnarray}
with
\begin{eqnarray}
   {\cal L}_{auxiliary} \ &=& \ - \,\,\frac{1}{8 \,G_S} \,\,
   \sum_a \,\,{( \,S^a \ + \ 4 \,G_S \,\,\bar{q} \,T^a \,q \,)}^2
   \nonumber \\
   &\,& \ - \,\,\frac{1}{8 \,G_S} \,\,
   \sum_a \,\,{( \,P^a \ + \ 4 \,G_S \,\,\bar{q} \,\,i \,\gamma_5 \,
   T^a \,q \,)}^2  \nonumber \\
   &\,& \ + \,\,\frac{1}{8 \,G_V} \,\,
   \sum_a \,\,{( \,V^{a \mu} \ + \ 4 \,G_V \,\,\bar{q} \,\,
   \gamma^\mu \,
   T^a \,q \,)}^2  \nonumber \\
   &\,& \ + \,\,\frac{1}{8 \,G_V} \,\,
   \sum_a \,\,{( \,A^{a \mu} \ + \ 4 \,G_V \,\,\bar{q} \,\,
   \gamma^\mu \,
   \gamma_5 \, T^a \,q \,)}^2  \,\, .
\end{eqnarray}
Here $S^a$, $P^a$, $V^{a \mu}$, and $A^{a \mu}$ are collective scalar,
pseudouscalar, vector and axial-vector fields. Defining the quantities
(henceforth, the summation symbol for the repeated flavor
index $a$ will be suppressed),
\begin{equation}
   S \ = \ S^a \,T^a, \ \ P \ = \ P^a \,T^a, \ \ V_\mu \ = \ 
   - \,i \,V_\mu^a \,T^a, \ \ A_\mu \ = \ - \,i \,A_\mu^a \,T^a,
\end{equation}
we can write the lagrangian as
\begin{eqnarray}
   {\cal L} \ \ &=& \ \ \bar{q} \,\,[ \,\,i \,\gamma^\mu \,( \,
   \partial_\mu \, + \, V_\mu \, + \, \gamma_5 \,A_\mu \,) \ - \ 
   ( \,S \, + \, i \,\gamma_5 \,P \,) \,] \,q \nonumber \\
   &\,& \ - \,\,\,\frac{1}{4 \,G_S} \,\,
   \mbox{tr} \,[ \,S^2 \, + \, P^2 \,]
   \ - \ \frac{1}{4 \,G_V} \,\,\mbox{tr} \,[ \,V_\mu^2 \, + \, 
   A_\mu^2 \,] \,\, .
\end{eqnarray}
Since there are no kinetic terms for the collective meson fields
$S^a$, $P^a$, $V^{a \mu}$, and $A^{a \mu}$, they are auxiliary at this
stage, and the lagrangian (2.5) is classically equivalent to the original
one (2.1).
It is now convenient to rewrite (2.5) using the left-right notation
as $R_\mu \,= \,V_\mu \, + \, A_\mu$, $L_\mu \, = \,V_\mu \, - \, A_\mu$,
and $M \, = \, S \, + \, i \,P$ :
\begin{eqnarray}
   {\cal L} \ \ &=& \ \ \bar{q} \,\,[ \,\,i \,\gamma^\mu \,( \,\partial_\mu
   \, + \, L_\mu \, + \,R_\mu \,) \ - \ 
   ( \,M^\dagger \,P_L \, + \,M \,P_R \,)
   \,\,] \,\,q \nonumber \\
   &\,& - \,\,\,\frac{1}{4 \,G_S} \,\,
   \mbox{tr} \,M \,M^\dagger
   \ - \ \frac{1}{8 \,G_V} \,\,\mbox{tr} \,[ \,L^2_\mu \ + \ R^2_\mu \,]
   \,\, ,
\end{eqnarray}
where $P_{R/L} = \frac{1}{2} \,(1 \pm \gamma_5)$ being chirality projection
operators. Customarily, the complex field $M$ is parameterized as
\begin{equation}
   M \ \ = \ \ \xi \,\,\Sigma \,\,\xi \,\, ,
\end{equation}
in terms of a hermitian matrix $\Sigma$ and a unitary matrix $\xi$, where
the latter is written as $\xi (x) = e^{i \,\pi (x) / f_\pi}$ in terms of
the Goldstone boson $\pi (x) = \pi^a (x) \,T^a$.
According to Bando, Kugo, and Yamawaki [6],
however, an arbitrary complex matrix can be expressed as a product of a
unitary matrix $U$ and a positive hermitian matrix $\tilde{H}$, so that
one may generally rewrite the complex matrix $M$ as
\begin{equation}
   M \ \ = \ \ U \,\,\tilde{H} \ \ = \ \ \xi_L^\dagger \,\,H \,\,
   \xi_R \,\, ,
\end{equation}
where the second equality is obtained by introducing the two unitary
matrices $\xi_L$ and $\xi_R$ :
\begin{eqnarray}
   U \ \ &=& \ \ \ \xi_L^\dagger \,\,\xi_R \,\, , \\
   \tilde{H} \ \ &=& \ \ \xi_R^\dagger \,\,H \,\, \xi_R \,\, .
\end{eqnarray}
As pointed out by them, the decomposition of $U$ into $\xi_L$ and $\xi_R$
in (2.9) is not unique and this arbitrariness is related to the appearance
of the hidden local symmetry ${U(n)}_V$ discussed below. The chiral
transformation law of $M$ follows from the invariance of
$\bar{\psi} \,M \,P_R \,\psi = {\bar{\psi}_L} \,M \,\psi_R$ :
\begin{equation}
   M \ \ \longrightarrow \ \ M' \ \ = \ \ g_L \,\,M \,\,g_R^\dagger \,\, .
\end{equation}
This , together with (2.8), (2.9), and (2.10), leads to the transformation
laws of $\xi_{L,R}$ and $H$ :
\begin{eqnarray}
   \xi_L (x) \ \ &\longrightarrow& \ \ \xi^\prime_L (x) \ \ = \ \ 
   h(x) \,\,\xi_L (x) \,\,g_L^\dagger  \,\, , \\
   \xi_R (x) \ \ &\longrightarrow& \ \ \xi^\prime_R (x) \ \ = \ \ 
   h(x) \,\,\xi_R (x) \,\,g_R^\dagger  \,\, , \\
   H (x) \ \ &\longrightarrow& \ \ H^\prime (x) \ \ = \ \ 
   h(x) \,\,H (x) \,\,h^\dagger (x) \, ,
   \ \ \ \ \ h(x) \ \in \ {[U(n)_V]}_{local} \,\, . \ \ \ 
\end{eqnarray}
It is possible to further enlarge the above hidden local symmetry to
${[U(n)_L \times U(n)_R]}_{local}$ by introducing another dynamical
variable $\xi_M (x)$ in such a way that [8,9]
\begin{equation}
   U (x) \ \ = \ \ \xi^\dagger_L (x) \,\,\xi_M (x) \,\, \xi_R (x) \,\, .
\end{equation}
The transformation properties of $\xi_{L,R}$ and $\xi_M$ under
$(g_L,g_R) \, \in \, {[U(n)_L \times U(n)_R]}_{global}$ and
$(h_L (x), h_R (x)) \, \in \, {[U(n)_L \times U(n)_R]}_{local}$ are given
by
\begin{eqnarray}
   \xi_{L,R} (x) \ \ &\longrightarrow& \ \ \xi^\prime_{L,R} (x) \ = \ 
   h_{L,R} (x) \,\,\xi_{L,R} \,\,g^\dagger_{L,R} \,\, , \\
   \xi_M (x) \ \ &\longrightarrow& \ \ \xi^\prime_M (x) \ = \ 
   h_L (x) \,\,\xi_M (x) \,\,h^\dagger_R (x) \,\, .
\end{eqnarray}
The redundant nature of the representation (2.15) can most clearly be seen
by introducing a parametrization as [9]
\begin{eqnarray}
   \xi_{L,R} (x) \ \ &=& \ \ e^{\pm \,i \,p(x) \,/ \,f_\pi} \,\cdot \,
   e^{i \,\sigma (x) \,/ \,f_\pi} \,\cdot \,
   e^{\mp \,i \,\pi (x) \,/ \,f_\pi} \,\, , \\
   \xi_M (x) \ \ &=& \ \ e^{2 \,i \,p(x) \,/ \,f_\pi} \,\, .
\end{eqnarray}
This reveals that the realization of the extended hidden local symmetry
is accomplished via the introduction of two kinds of ``compensating
fields'', $p(x) = p^a (x) \,T^a$ and $\sigma (x) = \sigma^a (x) \,T^a$,
which play the role of the gauge parameters to be absorbed into the
masses of the vector and axial-vector mesons in the unitary gauge such that
\begin{equation}
   p(x) \ \ = \ \ 0 \hspace{6mm} \mbox{or} \hspace{6mm}
   \xi_M (x) \ \ = \ \ 1 \,\, ,
\end{equation}
and further
\begin{equation}
   \sigma (x) \ \ = \ \ 0 \hspace{6mm} \mbox{or} \hspace{6mm}
   \xi^\dagger_L (x) \ = \ \xi_R (x) \ = \ \xi (\pi) \ = \ 
   e^{i \,\pi (x) \,/ \,f_\pi} \,\, ,
\end{equation}
respectively.

From now on, we adopt the most general representation (2.15) for
$U(x)$, while replacing the hermitian matrix $H (x)$ by
its vacuum expectation value $< \! H (x) \!> \,= m$ with $m$ being the
dynamical quark mass generated through
the spontaneous chiral symmetry breaking of the QCD vacuum. This latter
approximation is motivated by the fact that we are interested in an
effective meson action which does not contain physical scalar fields.
(This is just the standard motivation for considering nonlinear realization
of chiral symmetry [1-6].) The lagrangian (2.6) is then written as
\begin{eqnarray}
   {\cal L} \ &=& \ \bar{q} \,\,[ \,i \,\gamma^\mu \,( \,
   \partial_\mu \, + \, L_\mu \,P_L \, + \,R_\mu \,P_L \,) \, - \, m \,
   ( \,\xi^\dagger_R \,\xi^\dagger_M \,\xi_L \,P_L \, + \, 
   \xi^\dagger_L \,\xi_M \,\xi_R \,P_R \,) \,] \,q \,\, \nonumber \\
   &\,& \hspace{2mm} - \,\,\,\frac{1}{8 \,G_V} \,\,\mbox{tr} \,
   [ \,L^2_\mu \ + \ R^2_\mu \,] \,\, .
\end{eqnarray}
Here we have dropped an irrelevant constant term. It is convenient to
introduce new fermion variables via a field dependent chiral rotation
(Weinberg rotation) as
\begin{eqnarray}
   \chi_L (x) \ \ \equiv \ \ \xi_L (x) \,\,q_L (x) \, , \ \ \ 
   \chi_R (x) \ \ \equiv \ \ \xi_R (x) \,\,q_R (x) \,\, .
\end{eqnarray}
Then, we can rewrite the lagrangian as
\begin{eqnarray}
   {\cal L} \ \ &=& \ \ \chi \,\,[ \,i \,\gamma^\mu \,( \,\partial_\mu
   \ + \ \tilde{L}_\mu \,P_L \ + \ \tilde{R}_\mu \,P_R \,) \ - \ m \,\,
   ( \, \xi^\dagger_M \,P_L \ + \ \xi_M \, P_R \,) \,] \,\chi \nonumber \\
   &\,& \ - \,\,\,\frac{1}{8 \,G_V} \,\,[ \,\,
   \,\mbox{tr} \,{( \,D_L \,\xi_L \,\cdot \,\xi^\dagger_L \,)}^2 \ + \ 
   \,\mbox{tr} \,{( \,D_R \,\xi_R \,\cdot \,\xi^\dagger_R \,)}^2 \,\,]
   \,\, ,
\end{eqnarray}
where
\begin{eqnarray}
   D_L \,\,\xi_L \ \ &=& \ \ ( \,\partial_\mu \ \ + \ \ \tilde{L}_\mu \,)
   \,\,\xi_L \,\, , \\
   D_R \,\,\xi_R \ \ &=& \ \ ( \,\partial_\mu \ \ + \ \ \tilde{R}_\mu \,)
   \,\,\xi_R \,\, ,
\end{eqnarray}
with the following definition of the new vector and axial-vector fields :
\begin{eqnarray}
   \tilde{L}_\mu \ \ \equiv \ \ \xi_L \,\,( \,L_\mu \ \ + 
   \ \ \partial_\mu \,) \,\xi^\dagger_L \,\, , \\
   \tilde{R}_\mu \ \ \equiv \ \ \xi_R \,\,( \,R_\mu \ \ + 
   \ \ \partial_\mu \,) \,\xi^\dagger_R \,\, .
\end{eqnarray}
The transformation laws of $\tilde{L}_\mu$ and $\tilde{R}_\mu$ follows
from (2.27) and (2.28) :
\begin{eqnarray}
   \tilde{L}_\mu \ \ &\longrightarrow& \ \ \tilde{L}^\prime_\mu \ \ = \ \ 
   h_L (x) \,\,( \,\tilde{L}_\mu \ \ + \ \ \partial_\mu \,) 
   \,h^\dagger_L (x) \,\, , \\
   \tilde{R}_\mu \ \ &\longrightarrow& \ \ \tilde{R}^\prime_\mu \ \ = \ \ 
   h_R (x) \,\,( \,\tilde{R}_\mu \ \ + \ \ \partial_\mu \,) 
   \,h^\dagger_R (x) \,\, ,
\end{eqnarray}
which shows that they transform as gauge fields of the enlarged
hidden local symmetry ${[U(n)_L \times U(n)_R]}^{(HLS)}$.

Two remarks are in order here by following Bando, Kugo, and Yamawaki [6].
First, at the stage of lagrangian (2.6), the vacuum functional $Z$ is
given by
\begin{equation}
   Z \ \ = \ \ \int \,\,{\cal D} M \,\,{\cal D} M^\dagger \,\,
   {\cal D} L_\mu \,\,{\cal D} R_\mu \,\,\cdot \,Z_f \,\, ,
\end{equation}
with
\begin{equation}
   Z_f \ \ = \ \ \int \,\,{\cal D} q \,\,{\cal D} \bar{q} \,\,\,\,
   e^{\,i \,\int \,\, d^4 x \,\,{\cal L}} \,\, .
\end{equation}
When the variable $M$ is changed into $U$ and $\tilde{H}$ and further into
$\xi_L$, $\xi_R$, $\xi_M$ and $H$, the path integral measure
${\cal D} M \,{\cal D} M^\dagger$ becomes
\begin{eqnarray}
   {\cal D} M \,\,{\cal D} M^\dagger \ \ = \ \ {\cal D} U \,\,
   {\cal D} \tilde{H} \ \ = \ \ 
   {\cal D} \xi_L \,\,{\cal D} \xi_R \,\,{\cal D} \xi_M \,\,{\cal D} H \,\,
   \delta(\,\xi^\dagger_L \,- \,\xi_R \,) \,\,\delta(\,\xi_M \,- \,1 \,) 
   \,\, . \ \ 
\end{eqnarray}
Here the delta function parts $\delta(\,\xi^\dagger_L \,- \,\xi_R \,)$ and
$\delta(\,\xi_M \,- \,1 \,)$ are necessary, since otherwise the number of
degrees of freedom corresponding to the $U$ field would be tripled when
expressing $U$ in terms of $\xi_L$, $\xi_R$ and $\xi_M$.
This particular form of constraints corresponds to taking the unitary gauge
(2.20) and (2.21). However, it is clear that the existence of the hidden
local symmetry allows us to replace those constraints by more general gauge
fixing conditions. Once this fact is understood, we can concentrate on
the fermion part of the path integral $Z_f$, by leaving the gauge
fixing problem for later consideration.
Secondly, as is widely known, the fermion path integral measure
${\cal D} q \,{\cal D} \bar{q}$ is not invariant due to the
presence of chiral anomaly [35]. Instead, we have
\begin{equation}
   {\cal D} q \,\,{\cal D} \bar{q} \ \ = \ \ {(J_1)}^{N_c} \,\,
   {\cal D} \chi \,\,{\cal D} \chi^\dagger \,\, ,
\end{equation}
where ${(J_1)}^{N_c}$ is the Jacobian of the transformation (2.23).
If we carry out further change of fermion variables as
\begin{eqnarray}
   \varphi_L (x) \ \ &\equiv& \ \ \xi^\dagger_M (x) \,\,\chi_L (x) 
   \,\, , \\
   \varphi_R (x) \ \ &\equiv& \ \ \hspace{12mm} \chi_R (x) \,\, ,
\end{eqnarray}
then $Z_f$ can be written in the following three forms :
\begin{eqnarray}
   Z_f \ \ &=& \ \ \int \,\,{\cal D} q \,\,\,\,{\cal D} \bar{q} \,\,
   e^{\,i \,\int \,\,d^4 x \,\,\bar{q} \,\,D \,\,q} \nonumber \\
   &=& \ \ {J_1}^{N_c} \,\,\int \,\,
   {\cal D} \chi \,\,{\cal D} \bar{\chi} \,\,\,\,
   e^{\,i \,\int \,\,d^4 x \,\,\bar{\chi} \,\,\tilde{D} \,\,\chi}
   \nonumber \\
   &=& \ \ {J_1}^{N_c} \,\,{J_2}^{N_c} \,\,\int \,\,
   {\cal D} \varphi \,\,{\cal D} \bar{\varphi} \,\,\ \,
   e^{\,i \,\int \,\,d^4 x \,\,\bar{\varphi} \,\,
   \hat{D} \,\,\varphi} \,\, ,
\end{eqnarray}
with
\begin{eqnarray}
   D \, &=& \, i \,\gamma^\mu \,
   ( \,\partial_\mu \ + \ L_\mu \,\,P_L \ + \ R_\mu \,\,P_R \,) \ - \ 
   m \,( \,\xi^\dagger_R \,\,\xi^\dagger_M \,\,\xi_L \,\,P_L \ + \ 
   \xi^\dagger_L \,\,\xi_M \,\,\xi_R \,\,P_R \,) \,\, , \ \ \ \\
   \tilde{D} \, &=& \, i \,\gamma^\mu \,
   ( \,\partial_\mu \ + \ \tilde{L}_\mu \,\,P_L \ + \ 
   \tilde{R}_\mu \,\,P_R \,) \ - \ 
   m \,( \,\xi^\dagger_M \,\,P_L \ + \ 
   \xi_M \,\,P_R \,) \,\, ,\\
   \hat{D} \, &=& \, i \,\gamma^\mu \,
   ( \,\partial_\mu \ + \ \hat{L}_\mu \,\,P_L \ + \ 
   \hat{R}_\mu \,\,P_R \,) \ - \ m \,\, .
\end{eqnarray}
Here we have defined the new field variables as
\begin{eqnarray}
   \hat{L}_\mu \ \ &\equiv& \ \ \xi^\dagger_M \,\,( \,\tilde{L}_\mu
   \ \ + \ \ \partial_\mu \,) \,\,\xi_M \,\, , \\
   \hat{R}_\mu \ \ &\equiv& \ \ \tilde{R}_\mu \,\, ,
\end{eqnarray}
whereas ${J_2}^{N_c}$ in (2.45) is the Jacobian of the transformation
(2.35) and (2.36).
Formally carrying out the fermion path integral, we then obtain
\begin{eqnarray}
   Z_f \ \ &=& \ \ {(\,\mbox{det} \,\,D)}^{N_c} \\
   &=& \ \ {(J_1)}^{N_c} \,\,{(\,\mbox{det} \,\,\tilde{D})}^{N_c} \\
   &=& \ \ {(J_1)}^{N_c} \,\,{(J_2)}^{N_c} \,\,
   {(\,\mbox{det} \,\,\hat{D})}^{N_c} \,\, .
\end{eqnarray}
As is well known, the real part of $\log \,{(\mbox{det} \,D)}^{N_c}$
contributes to the non-anomalous (intrinsic-parity conserving) part of
effective action, whereas the imaginary part of it gives the anomalous
(intrinsic-parity violating) part [18,19].
(Here and hereafter, we frequently
use the terminology as above in the Euclidean formulation of the path
integral for convenience, in spite that we are working in the Minkowski
formulation.)
We also know that the modulus of the quark determinant is chiral
gauge invariant [18,19], i.e.
\begin{equation}
   | \,\mbox{det} \,\,D \,| \ \ = \ \ | \,\mbox{det} \,\,\tilde{D} \,|
   \ \ = \ \ | \,\mbox{det} \,\,\hat{D} \,| \,\, .
\end{equation}
This part contains divergences, which must be removed by some
regularization procedure. Here we adopt the proper-time regularization
scheme with some intrinsic cutoff $\Lambda$ :
\begin{eqnarray}
   N_c \,\,\log \,\,| \,\mbox{det} \,\,D \,| \ \ &=& \ \ 
   \frac{N_c}{2} \,\,\,\mbox{Tr}^\prime \,\,\log \,\,D^\dagger \,\,D
   \nonumber \\
   &\longrightarrow& \ \ - \,\,\,
   \frac{N_c}{2} \,\,\int_{1 \,/ \,\Lambda^2}^\infty
   \,\,\frac{d \tau}{\tau} \,\,\mbox{Tr}^\prime \,\,
   e^{- \,\tau \,\,D^\dagger \,\,D} \,\, .
\end{eqnarray}
Here $\mbox{Tr} \,= \,\int \,\,d^4 x \,\,\mbox{tr}$ and a prime on it
indicates that a trace over Dirac indices is included.
$D^\dagger \,D$ in the above equation can naturally be replaced by either
of $\tilde{D}^\dagger \,\tilde{D}$ or $\hat{D}^\dagger \,\hat{D}$.
It is also clear from (2.40) that $\mbox{det} \,\hat{D}$ has no imaginary
part or it is chiral gauge invariant, so that the imaginary part $\Delta$
of $\log \,\mbox{det} \,D$ comes from the two Jacobians $J_1$ and $J_2$,
leading to the following expression :
\begin{equation}
   \log \,{( \,\mbox{det} \,D \,)}^{N_c} \ \ = \ \ 
   N_c \,\,\log \,| \,\mbox{det} \, D \,| \ \ + \ \ i \,\Delta \,\, ,
\end{equation}
with
\begin{equation}
   \Delta \ \ = \ \ N_c \,\,\,\mbox{Im} \,\,( \,\log \,J_1 \ + \ 
   \log \,J_2 \,) \,\, .
\end{equation}
The real and imaginary parts of the effective meson action will be
discussed separately in the following two subsections.

\vspace{6mm}
\subsection{Non-anomalous effective action}

\ \ \ \ \ \ Since our main concern in this paper is the anomalous part
of the action,
we give here only a brief survey of the nonanomalous part of it, which
can be obtained by using the standard derivative expansion method [18-21].
As already pointed out, it is immaterial which form of Dirac operator,
i.e. the original one $D$, or the chirally rotated ones $\tilde{D}$ or
$\hat{D}$, is used in this evaluation (at least assuming infinite
summation of the gradient expansion). It is a matter of representation.
Here we adopt the form $\tilde{D}$, since we want to interpret
$\tilde{L}_\mu$ and $\tilde{R}_\mu$ as gauge bosons of the hidden local
$U(n)_L \times U(n)_R$ symmetry. (Remember the transformation laws
(2.29) and (2.30) of ${\tilde{\cal L}}_\mu$ and ${\tilde{\cal R}}_\mu$.)
As follows is the outline of the necessary manipulation [20,21].
First, truncate the derivative expansion at terms of
second order. Second, introduce the coupling constant $g_V$ by the
relation
\begin{equation}
   g_V \ \ = \ \ {\{ \,\frac{2}{3} \,\,\frac{N_c}{{(4 \,\pi)}^2} \,\,
   \Gamma \,( \,0, \frac{m^2}{\Lambda^2} \,) \,\} }^{- \,1 / 2} \,\, ,
\end{equation}
where $\Gamma (\alpha, x)$ is the incomplete gamma function defined by
$\Gamma (\alpha, x) \,= \,\int_x^\infty \,dt \,e^{- \,t} \,t^{\alpha -1}$.
Then, examining the coefficients of the bilinear terms of the field
variables in the resultant effective lagrangian, we find the following
relations :
\begin{equation}
   M_V^2 \ \ = \ \ \frac{g^2_V}{4 \,G_V} \, , \ \ \ \ 
   M^2_A \ \ = \ \ M^2_V \ + \ 6 \,\,m^2 \, , \ \ \ \ 
   f^2_\pi \ \ = \ \ \frac{1}{4 \,G_V} \,\,
   ( \,1 \ - \ \frac{M^2_V}{M^2_A} \,) \,\, .
\end{equation}
Here $M_V$ and $M_A$ are respectively the masses of the vector and axial-vector mesons, with $m$ being the dynamical quark mass [16-22].
Finally, defining the
parameter $a$ by the equation [20,21]
\begin{equation}
   a \ \ = \ \ {( \,1 \ - \ \frac{M^2_V}{M^2_A} \,)}^{- \,1} \,\, ,
\end{equation}
we are led to an effective meson lagrangian of the following form :
\begin{eqnarray}
   {{\Gamma}}^{(n)} \ \ \ = \ \ \ \int \,\,\, d^4 x \,\,\,
   {\cal L}^{(n)} \,\, ,
\end{eqnarray}
with
\begin{eqnarray}
   {\cal L}^{(n)} \ \ &=& \ \ \frac{1}{4 \,g^2_V} \,\,\mbox{tr} \,\,
   [ \,{\tilde{L}}^2_{\mu \nu} \ + \ {\tilde{R}}^2_{\mu \nu} \,]
   \ - \ \frac{a}{a - 1} \,\,\frac{f^2_\pi}{4} \,\,\mbox{tr} \,\,
   {[ \,D_\mu \,\xi_M \,\cdot \,\xi^\dagger_M \,]}^2 \nonumber \\
   &\,& \!\!\! - \,\,\,\,\frac{1}{2} \,\,a \,\,f^2_\pi \,\,\mbox{tr} \,\,
   {[ \,D_\mu \,\xi_L \,\cdot \,\xi^\dagger_L \,]}^2 \ \ - \ \ 
   \frac{1}{2} \,\,a \,\,f^2_\pi \,\,\mbox{tr} \,\,
   {[ \,D_\mu \,\xi_R \,\cdot \,\xi^\dagger_R \,]}^2 \,\, .
\end{eqnarray}
Here
\begin{eqnarray}
   {\tilde{L}}_{\mu \nu} \ \ &=& \ \ \partial_\mu \,\,{\tilde{L}}_\nu
   \ - \ \partial_\nu \,\,{\tilde{L}}_\mu \ + \ 
   [ \,{\tilde{L}}_\mu , \,{\tilde{L}}_\nu \,] \,\, , \\
   {\tilde{R}}_{\mu \nu} \ \ &=& \ \ \partial_\mu \,\,{\tilde{R}}_\nu
   \ - \ \partial_\nu \,\,{\tilde{R}}_\mu \ + \ 
   [ \,{\tilde{R}}_\mu , \,{\tilde{R}}_\nu \,] \,\, ,
\end{eqnarray}
and
\begin{eqnarray}
   D_\mu \,\,\xi_L \ \ &=& \ \ \partial_\mu \,\,\xi_L \ + \ 
   {\tilde{L}}_\mu \,\,\xi_L \,\, , \\
   D_\mu \,\,\xi_R \ \ &=& \ \ \partial_\mu \,\,\xi_R \ + \ 
   {\tilde{R}}_\mu \,\,\xi_R \,\, , \\
   D_\mu \,\,\xi_M \ \ &=& \ \ \partial_\mu \,\,\xi_M \ + \ 
   {\tilde{L}}_\mu \,\,\xi_M \ - \ \xi_M \,\,{\tilde R}_\mu \,\, .
\end{eqnarray}
The lagrangian above may be compared with that of Bando et al. derived
from the symmetry principle of enlarged hidden local symmetry [9] :
\begin{equation}
   {\cal L} \ \ = \ \ \frac{1}{4 \,g^2_V} \,\,\mbox{tr} \,\,
   [ \,{\tilde{L}}^2_{\mu \nu} \ + \ {\tilde{R}}^2_{\mu \nu} \,] \ + \ 
   a' \,\,{\cal L}_V \ + \ b' \,\,{\cal L}_A \ + \ c' \,\,L_M \ + \ 
   d' \,\,{\cal L}_\pi \,\, ,
\end{equation}
where
\begin{eqnarray}
   {\cal L}_V \ \ &=& \ \ - \,\,\,\frac{f^2_\pi}{4} \,\,\mbox{tr} \,\,
   {[ \,D_\mu \,\xi_L \,\cdot \,\xi^\dagger_L \ + \ \xi_M \,\,D_\mu \,\xi_R
   \,\cdot \,\xi^\dagger_R \,\,\xi^\dagger_M \,]}^2 \,\, , \\
   {\cal L}_A \ \ &=& \ \ - \,\,\,\frac{f^2_\pi}{4} \,\,\mbox{tr} \,\,
   {[ \,D_\mu \,\xi_L \,\cdot \,\xi^\dagger_L \ - \ \xi_M \,\,D_\mu \,\xi_R
   \,\cdot \,\xi^\dagger_R \,\,\xi^\dagger_M \,]}^2 \,\, , \\
   {\cal L}_M \ \ &=& \ \ - \,\,\,\frac{f^2_\pi}{4} \,\,\mbox{tr} \,\,
   {[ \,D_\mu \,\xi_M \,\cdot \,\xi^\dagger_M \,]}^2 \,\, , \\
   {\cal L}_\pi \ \ &=& \ \ - \,\,\,\frac{f^2_\pi}{4} \,\,\mbox{tr} \,\,
   {[ \,D_\mu \,\xi_L \,\cdot \,\xi^\dagger_L \ - \ \xi_M \,\,D_\mu \,\xi_R
   \,\cdot \,\xi^\dagger_R \,\,\xi^\dagger_M  \ - \ 
   D_\mu \,\xi_M \,\cdot \,\xi^\dagger_M \,]}^2 \,\, .
\end{eqnarray}
It is easy to see that their lagrangian (2.60) coincides with our
lagrangian (2.54), apart from the fact that the arbitrary constants
$a'$, $b'$, $c'$,
$d'$ of theirs are respectively constrained such that
$a' = b' = a$, $c' = a / (a -1)$, $d' = 0$ in our effective lagrangian,
which has been derived from a specific underlying lagrangian at the
quark level. Note that, from our
derivation here, it is self-evident that the equality of
$a'$ and $b'$ is simply a reflection of the chiral symmetry satisfied
by the original NJL lagrangian [21].

\vspace{4mm}
\subsection{Anomalous effective action}

\ \ \ \ \ \ Now we turn to the discussion of the intrinsic-parity
violating part of
the effective action, which is of our primary concern in this paper :
\begin{eqnarray}
   \Gamma^{(a)} \ \ &=& \ \ N_c \,\,\mbox{Im} \,\log \,
   \mbox{det} \,\,D \nonumber \\
   &=& \ \ - \,\,i \,\,N_c \,\,( \,\log \,J_1 \ + \ \log \,J_2 \,) \,\, .
\end{eqnarray}
Here use has been made of the fact that the Jacobians $J_1$ and $J_2$ are
pure phase [18,19].
It is easier to first evaluate the response of the Jacobians
under small chiral variations of the relevant fields, rather than
calculating the Jacobians directly. The answer depends on the
regularization scheme. Since the presence of anomaly does not allow
us to maintain
all the symmetries of the classical level lagrangian, a general question
is what symmetries should be kept unbroken after regularization.
At the classical level, our lagrangian has the $U(n)_L \times U(n)_R$
global symmetry and the $U(n)_L \times U(n)_R$ hidden local symmetries.
One might then wonder which symmetry should be kept unbroken
in our particular problem.
Fortunately, in the absence of the coupling with the external
electroweak fields, the answer is quite simple. In fact, we can retain
both of these symmetries despite the presence of the anomaly.
(As a matter of course, what is truly interesting from the physical
viewpoint is the realistic case with the electroweak couplings.
We discuss this physically interesting problem separately in the next
section.) 

Let us now construct the anomalous action explicitly, which has
${[U(n)_L \times U(n)_R]}^{(ext)}_{global}$ and
${[U(n)_L \times U(n)_R]}^{(HLS)}$ symmetries. The most elegant way to
carry out this program is to use the differential geometric
method [36-39]. (Here we closely follow the treatment by Petersen [39].)
For convenience sake, let us consider space-time dependent chiral
transformations $(g_L (x), g_R (x)) \in 
{[U(n)_L \times U(n)_R]}^{(ext)}_{local}$, which generalizes the global
chiral transformation. Its infinitesimal form is given as
\begin{eqnarray}
   g_L (x) \ \ &=& \ \ e^{- \,\theta_L (x)} \ \ \simeq \ \ 
   1 \ - \ \theta_L (x) \,\, , \\
   g_R (x) \ \ &=& \ \ e^{- \,\theta_R (x)} \ \ \simeq \ \ 
   1 \ - \ \theta_R (x) \,\, ,
\end{eqnarray}
with the property $\theta^\dagger_L = - \,\theta_L$,
$\theta^\dagger_R = - \,\theta_R$.
We list below the transformation properties
of the relevant fields under this transformation :
\begin{eqnarray}
   \xi_L (x) \ &\longrightarrow& \ \xi_L (x) \,\,g^\dagger_L (x) \, ,
   \hspace{28mm} \mbox{or} \hspace{5mm}
   \delta^{(ext)} \,\,\xi_L \ \ = \ \ \xi_L \,\,\theta_L \,\, , \\
   \xi_R (x) \ &\longrightarrow& \ \xi_R (x) \,\,g^\dagger_R (x) \, ,
   \hspace{27mm} \mbox{or} \hspace{5mm}
   \delta^{(ext)} \,\,\xi_R \ \ = \ \ \xi_R \,\,\theta_R \,\, , \\
   \xi_M (x) \ &\longrightarrow& \ \xi_M (x) \, ,
   \hspace{38mm} \mbox{or} \hspace{5mm}
   \delta^{(ext)} \,\,\xi_M \ \ = \ \ 0 \,\, , \\
   L(x) \ &\longrightarrow& \ g_L (x) \,\,
   (\,L (x) \ + \ d \,) g^\dagger_L (x) \, ,
   \hspace{5mm} \mbox{or} \hspace{4mm}
   \delta^{(ext)} \,\,L \ \ = \ \ d \,\theta_L \ + \ [ \,L, \theta_L \,]
   \,\, , \\
   R(x) \ &\longrightarrow& \ g_R (x) \,\,
   (\,R (x) \ + \ d \,) g^\dagger_R (x) \, ,
   \hspace{5mm} \mbox{or} \hspace{4mm}
   \delta^{(ext)} \,\,R \ \ = \ \ d \,\theta_R \ + \ [ \,R, \theta_R \,]
   \,\, , \ \ \ \\
   \tilde{L} (x) \ &\longrightarrow& \ \tilde{L} (x) \, ,
   \hspace{41mm} \mbox{or} \hspace{4mm}
   \delta^{(ext)} \,\,\tilde{L} \ \ = \ \ 0 \,\, , \\
   \tilde{R} (x) \ &\longrightarrow& \ \tilde{R} (x) \, ,
   \hspace{41mm} \mbox{or} \hspace{4mm}
   \delta^{(ext)} \,\,\tilde{R} \ \ = \ \ 0 \,\, , \\
   \hat{L} (x) \ &\longrightarrow& \ \hat{L} (x) \, ,
   \hspace{41mm} \mbox{or} \hspace{4mm}
   \delta^{(ext)} \,\,\hat{L} \ \ = \ \ 0 \,\, , \\
   \hat{R} (x) \ &\longrightarrow& \ \hat{R} (x) \, ,
   \hspace{41mm} \mbox{or} \hspace{4mm}
   \delta^{(ext)} \,\,\hat{R} \ \ = \ \ 0 \,\, .
\end{eqnarray}
Here and hereafter, we use the notation of differential form with the
definition $L = L_\mu \,d x^\mu$, $R = R_\mu \,d x^\mu$ and
$d = d x^\mu \,\partial_\mu$ etc.

Similarly, under the transformation $(h_L (x), h_R (x)) \in 
{[U(n)_L \times U(n)_R]}^{(HLS)}$ with
\begin{eqnarray}
   h_L (x) \ \ &=& \ \ e^{- \,\epsilon_L (x)} \ \ \simeq \ \ 
   1 \ - \ \epsilon_L (x) \,\, , \\
   h_R (x) \ \ &=& \ \ e^{- \,\epsilon_R (x)} \ \ \simeq \ \ 
   1 \ - \ \epsilon_R (x) \,\, ,
\end{eqnarray}
the relevant fields transform as
\begin{eqnarray}
   \xi_L (x) \, &\longrightarrow& \, h_L (x) \,\,\xi_L (x) \, ,
   \hspace{26mm} \mbox{or} \hspace{4mm}
   \delta^{(HLS)} \,\,\xi_L \ = \ - \,\epsilon_L \,\,\xi_L 
   \,\, , \\
   \xi_R (x) \, &\longrightarrow& \, h_R (x) \,\,\xi_R (x) \, ,
   \hspace{26mm} \mbox{or} \hspace{4mm}
   \delta^{(HLS)} \,\,\xi_R \ = \ - \,\epsilon_R \,\,\xi_R
   \,\, , \\
   \xi_M (x) \, &\longrightarrow& \, h_L (x) \,\,\xi_M (x) \,\,
   h^\dagger_R (x) \, ,
   \hspace{13mm} \mbox{or} \hspace{4mm}
   \delta^{(HLS)} \,\,\xi_M \ = \ \xi_M \,\,\epsilon_R \ - \ 
   \epsilon_L \,\,\xi_M \,\, , \\
   L (x) \, &\longrightarrow& \, L(x) \, ,
   \hspace{41mm} \mbox{or} \hspace{4mm}
   \delta^{(HLS)} L(x) \ = \ 0 \,\, , \\
   R (x) \, &\longrightarrow& \, R(x) \, ,
   \hspace{41mm} \mbox{or} \hspace{4mm}
   \delta^{(HLS)} R(x) \ = \ 0 \,\, , \\
   \tilde{L} (x) \, &\longrightarrow& \, h_L (x) \,\,
   ( \,\tilde{L} (x) \ + \ d \,) \,\,h^\dagger_L (x) \, ,
   \hspace{4mm} \mbox{or} \hspace{4mm}
   \delta^{(HLS)} \tilde{L} \ = \ d \,\epsilon_L \ + \ 
   [ \,\tilde{L}, \epsilon_L \,] \,\, , \\
   \tilde{R} (x) \, &\longrightarrow& \, h_R (x) \,\,
   ( \,\tilde{R} (x) \ + \ d \,) \,\,h^\dagger_R (x) \, ,
   \hspace{3mm} \mbox{or} \hspace{4mm}
   \delta^{(HLS)} \tilde{R} \ = \ d \,\epsilon_R \ + \ 
   [ \,\tilde{R}, \epsilon_R \,] \,\, , \\
   \hat{L} (x) \, &\longrightarrow& \, h_R (x) \,\,
   ( \,\hat{L} (x) \ + \ d \,) \,\,h^\dagger_R (x) \, ,
   \hspace{3mm} \mbox{or} \hspace{4mm}
   \delta^{(HLS)} \hat{L} \ = \ d \,\epsilon_R \ + \ 
   [ \,\hat{L}, \epsilon_R \,] \,\, , \\
   \hat{R} (x) \, &\longrightarrow& \, h_R (x) \,\,
   ( \,\hat{R} (x) \ + \ d \,) \,\,h^\dagger_R (x) \, ,
   \hspace{3mm} \mbox{or} \hspace{4mm}
   \delta^{(HLS)} \hat{R} \ = \ d \,\epsilon_R \ + \ 
   [ \,\hat{R}, \epsilon_R \,] \,\, .
\end{eqnarray}
The basic differential geometric object, which plays important roles in the
following construction is the so-called Chern-Simons secondary form
defined as
\begin{equation}
   \omega^0_{2n + 1} \,(A_L, A_R) \ \ = \ \ (n + 1) \,\,\int_{-1}^1 \,\,
   dt \,\,\mbox{tr} \,\,[ \,\dot{A} (t) \,\,F^n (t) \,] \,\, ,
\end{equation}
where $\dot{A} (t) = (d / dt) \,A(t)$ with
\begin{eqnarray}
   A(t) \ \ &=& \ \ \frac{1}{2} \,\,( \,A_L \, + \, A_R \,) \, - \, 
   \frac{1}{2} \,\,( \,A_L \, - \, A_R \,) \,\,t \,\, , \\
   F(t) \ \ &=& \ \ d \,A(t) \ + \ A^2 (t) \,\, ,
\end{eqnarray}
and $n = D / 2$ with $D$ being the space-time dimension. (We are of course
interested in the case with $D = 4$.) The special choice of the above
integral path in the field space (the straight line connecting $A_L$ and
$A_R$) dictates that $\omega^0_5 (A_L,A_R)$ is invariant under the
vector-type gauge transformation of $A_L$ and $A_R$, or that
it gives the so-called Bardeen anomaly. We also need the following quantity
defined by the above $\omega^0_5 (A_L,A_R)$ as
\begin{equation}
   {\hat{\omega}}^0_5 (A_L,A_R) \ \ \equiv \ \ \omega^0_5 (A_L,0) \ - \ 
   \omega^0_5 (A_R,0) \,\, .
\end{equation}
This quantity is not invariant under either of the vector-type or
axial-vector-type gauge transformation, whereas it gives the so-called
left-right symmetric form of anomaly. An important observation is that
the difference of $\omega^0_5 (A_L,A_R)$ and ${\hat{\omega}}^0_5 (A_L,A_R)$
is an exact form. In fact, it is known that one can write as [39]
\begin{equation}
   \omega^0_5 (A_L,A_R) \ - \ {\hat{\omega}}^0_5 (A_L,A_R) \ \ = \ \ 
   d \,\rho_4 \,(0,A_L,A_R) \,\, ,
\end{equation}
where
\begin{eqnarray}
   \rho_{2n} (A_0,A_1,A_2) \, = \, - \,\,(n+1) \,\,\sum_{p=0}^{n-1} \,\,
   \int_0^1 \,\,ds \,\,\int_0^{1-s} \, dt \,\,\mbox{tr} \,\,
   \{ \,A_2 \,F^p (s,t) \,A_1 \,F^{n-p-1} (s,t) \,\} \,\, , \ \ \ \ \ 
\end{eqnarray}
with
\begin{eqnarray}
   A(s,t) \ &=& \ A_0 \ + \ s \,\,A_1 \ + \ t \,\,A_2 \,\, , \\
   F(s,t) \ &=& \ d \,A(s,t) \ + \ A^2 (s,t) \,\, .
\end{eqnarray}
Specializing to the case of $D=4$ with $A_0 = 0$, $A_1 = A_L$, $A_2 = A_R$,
we obtain
\begin{eqnarray}
   \rho_4 \,(\,0, \,A_L, \,A_R \,) \ &=& \ 
   \frac{1}{2} \,\,\mbox{tr} \,\,[ \,\,
   (\,A_L \,A_R \, - \,A_R \,A_L \,) \,\,
   ( \,F_L \,+ \,F_R \,) \nonumber \\
   &\,& \! - \,\,\,A^3_L \,A_R \,+ \,A^3_R \,A_L \,+ \,
   \frac{1}{2} \,\,A_L \,A_R \,A_L \,A_R \,\,] \,\, .
\end{eqnarray}
(This is what we need later for writing down the explicit form of anomalous
action with the required symmetry.)
Using these differential geometric objects, the anomalous part of the
action, which satisfies the requirements above, i.e. the invariance under
${[U(n)_L \times U(n)_R]}^{(ext)}_{global}$ and
${[U(n)_L \times U(n)_R]}^{(HLS)}$, can easily be written down as
\begin{eqnarray}
   \Gamma^{(a)} \ \ &=& \ \ - \,\,\,i \,\,N_c \,\,( \,\log \,J_1 \ + \ 
   \log \,J_2 \,) \,\, ,\nonumber \\
   &=& \ \ c' \,\,\,\int_{B^5} \,\,\,\{ \,
   [ \,{\hat{\omega}}^0_5 (L,R) \ - \ 
   \omega^0_5 (\tilde{L}, \tilde{R}) \,] \ \ + \ \ 
   [ \,\omega^0_5 (\tilde{L}, \tilde{R}) \ - \ 
   \omega^0_5 (\hat{L}, \hat{R}) \,] \,\} \,\, ,\nonumber \\
   &=& \ \ c' \,\,\,\int_{B^5} \,\,\,[ \,{\hat{\omega}}^0_5 (L,R)
   \ \ - \ \ 
   \omega^0_5 (\hat{L}, \hat{R}) \,] \,\, ,
\end{eqnarray}
where $c' = i \,(N_c / \, 24 \,\pi^2)$, while $B^5$ is a
five-dimensional
manifold with the four dimensional space-time as its boundary.
Here $[\,{\hat{\omega}}^0_5 (L,R) \ - \ \omega^0_5 
(\tilde{L}, \tilde{R}) \,]$ part corresponds to the contribution from
the Jacobian $J_1$, while $[\,\omega^0_5 (\tilde{L}, \tilde{R}) \ - \ 
\omega^0_5 (\hat{L}, \hat{R}) \,]$
from $J_2$.

The fact that the above $\Gamma^{(a)}$ has the required properties
can be convinced as follows. First notice that, under the
${[U(n)_L \times U(n)_R]}^{(ext)}_{local}$ transformation, $L$ and $R$
transform according to (2.71) and (2.72),
whereas $\hat{L}$ and $\hat{R}$ are
absolutely intact under the same transformation. One then sees that
only the $\omega^0_5 (L,R)$ part in the last equation of (2.89)
changes under this
transformation, which just gives the left-right symmetric form of anomaly
by construction. This means that $\Gamma_{(a)}$ is invariant under the
global chiral transformation, which is a special case of
${[U(n)_L \times U(n)_R]}^{(ext)}_{local}$.

On the other hand, under the ${[U(n)_L \times U(n)_R]}^{(HLS)}$
transformation, $L$ and $R$ do not change, while $\hat{L}$ and $\hat{R}$
transform according to (2.86) and (2.87).
Here, a crucial observation is that
$\hat{L}$ and $\hat{R}$ transform exactly in the same manner.
In other words, for the fields $\hat{L}$ and $\hat{R}$ introduced
by (2.41) and (2.42),
only the vector-type transformation is induced by arbitrary
${[U(n)_L \times U(n)_R]}^{(HLS)}$ transformation .
Since $\omega^0_5 (\hat{L},\hat{R})$
is vector gauge invariant by construction, one then concludes that
$\Gamma^{(a)}$ is invariant under ${[U(n)_L \times U(n)_R]}^{(HLS)}$,
which insists that there is no anomaly in the enlarged hidden local
symmetry.

Although the promised effective action has already been given in a
formal sense, it is desirable to write down its explicit form.
This is especially so, because our goal is to express the effective action
in terms of $\xi_L$, $\xi_R$, $\xi_M$, $\tilde{L}$ and $\tilde{R}$, which
are the dynamical variables of the enlarged hidden local symmetry
scheme (or representation). It can be achieved by considering the integral
path in the field space as illustrated in fig.1(a). This integral path is
obtained by combining the two paths shown in fig.2(b), which are
respectively related to the contribution from the Jacobians
$J_1$ and $J_2$. From fig.1(a), we obtain
\begin{eqnarray}
   &\,& \omega^0_5 (L,0) \ + \ \omega^0_5 (0,R) \ + \ 
   \omega^0_5 (R,\xi^\dagger_R \,d \,\xi_R) \ + \ 
   \omega^0_5 (\xi^\dagger_R \,d \,\xi_R,0) \nonumber \\
   &+& \omega^0_5 (0,\tilde{R}) \ + \ \omega^0_5 (\hat{R},\hat{L}) \ + \ 
   \omega^0_5 (\hat{L},0) \ + \ \omega^0_5 (0,\xi_M \,d \,\xi^\dagger_M)
   \nonumber \\
   &+& \omega^0_5 (\xi_M \,d \,\xi^\dagger_M, \tilde{L}) \ + \ 
   \omega^0_5 (\tilde{L},0) \ + \ \omega^0_5 (0,\xi^\dagger_L \,d \xi_L)
   \ + \ \omega^0_5 (\xi^\dagger_L \,d \xi_L,L) \nonumber \\
   &=& d \,\,\{ \, \rho_4 (0,R,\xi^\dagger_R \,d \,\xi_R) \, + \, 
   \rho_4 (0,\hat{R},\hat{L}) \, + \, 
   \rho_4 (0,\xi_M \,d \,\xi^\dagger_M,\tilde{L}) \, + \, 
   \rho_4 (0,\xi^\dagger_L \,d \,\xi_L, L) \,\} \,\, . \ \ \ 
\end{eqnarray}
Now we can make use of the vector gauge invariant property of
$\omega^0_5 (A_L,A_R)$. For instance, we can show that
\begin{eqnarray}
   \omega^0_5 (R,\xi^\dagger_R \,d \,\xi_R) \ \ &=& \ \ 
   T \,(\,\xi_R, \xi_R) \,\,\omega^0_5 \,\,
   (R,\xi^\dagger_R \,d \,\xi_R)
   \nonumber \\
   &=& \ \ \omega^0_5 (\,\xi_R \,(\,R \ + \ d \,) \,\xi^\dagger_R, 0)
   \ \ = \ \ \omega^0_5 (\,\tilde{R}, 0) \,\, ,
\end{eqnarray}
where $T(g_L, g_R)$ defines the action of the element
$g = g_L \,P_L \,\,+ \,\,g_R \,P_L \in U(n)_L \times U(n)_R$. Then,
by utilizing the antisymmetry of $\omega^0_5 (A_L, A_R)$ with respect
to the interchange of $A_L$ and $A_R$, i.e. $\omega^0_5 (A_L, A_R) \, = \,
- \,\omega^0_5 (A_R, A_L)$, we find
\begin{eqnarray}
   \omega^0_5 \,(R,\xi^\dagger_R \,d \,\xi_R) \ + \ \omega^0_5 \,
   (0, \tilde{R}) \ \ = \ \ 0 \,\, .
\end{eqnarray}
Similarly, it is easy to verify that
\begin{eqnarray}
   \omega^0_5 \,(\hat{L}, 0) \ + \ \omega^0_5 \,
   (\xi_M \,d \,\xi^\dagger_M, \tilde{L}) \ \ &=& \ \ 0 \,\, , \\
   \omega^0_5 \,(\tilde{L}, 0) \ \,\,+ \ \,\,\omega^0_5 \,
   (\xi^\dagger_L \,d \,\xi_L, L) \ \ &=& \ \ 0 \,\, .
\end{eqnarray}
Using these equalities, we obtain from (2.98)
\begin{eqnarray}
   &\,& {\hat{\omega}}^0_5 \,(L,R) \ - \ \omega^0_5 \,(\hat{L},\hat{R})
   \nonumber \\
   &=& \omega^0_5 \,(0,\xi^\dagger_R \,d \,\xi_R) \ - \ 
   \omega^0_5 \,(0,\xi^\dagger_L \,d \,\xi_L) \ - \ 
   \omega^0_5 \,(0,\xi_M \,d \,\xi^\dagger_M) \nonumber \\
   &+& d \,\,\{ \,\rho_4 \,(0, R, \xi^\dagger_R \,d \,\xi_R) \ + \ 
   \rho_4 \,(0, \xi^\dagger_L \,d \,\xi_L, L) \ + \ 
   \rho_4 \,(0, \xi_M \,d \,\xi^\dagger_M, \tilde{L}) \ - \ 
   \rho_4 \,(0, \hat{L}, \hat{R}) \,\} \,\, , \ \ 
\end{eqnarray}
where use has been made of the relation (2.91). Upon integration over the
5-dimensional manifold with space-time boundary, we then arrive at
\begin{eqnarray}
   \Gamma^{(a)} \ \ &\equiv& \ \ c' \,\,\,\int_{B^5} \,\,[ \,\,
   {\hat{\omega}}^0_5 \,(L,R) \ - \ \omega^0_5 (\hat{L},\hat{R}) \,]
   \nonumber \\
   &=& \ \ c' \,\,\,\int_{B^5} \,\,[ \,\,
   \omega^0_5 \,( 0, \xi^\dagger_R \,d \xi_R) \ - \ 
   \omega^0_5 \,( 0, \xi^\dagger_L \,d \xi_L) \ - \ 
   \omega^0_5 \,( 0, \xi_M \,d \,\xi^\dagger_M) \,] \nonumber \\
   &+& \ \ c' \,\,\,\int_{S^4} \,\,[ \,\,
   \rho_4 \,(0, R, \xi^\dagger_R \,d \,\xi_R) \ + \ 
   \rho_4 \,(0, \xi^\dagger_L \,d \,\xi_L, L) \nonumber \\
   &\,& \hspace{14mm} + \ \ 
   \rho_4 \,(0, \xi_M \,d \,\xi^\dagger_M, \tilde{L}) \ - \ 
   \rho_4 \,(0, \hat{L}, \hat{R}) \,\,\,] \,\, ,
\end{eqnarray}
or equivalently
\begin{eqnarray}
   \Gamma^{(a)} \ 
   &=& \ c' \,\,\,\int_{B^5} \,\,[ \,
   \omega^0_5 \,( 0, \xi^\dagger_R \,d \xi_R) \ - \ 
   \omega^0_5 \,( 0, \xi^\dagger_L \,d \xi_L) \ - \ 
   \omega^0_5 \,( 0, \xi_M \,d \,\xi^\dagger_M) \,] \nonumber \\
   &+& \ c' \,\,\,\int_{S^4} \,\,[ \,
   \rho_4 \,(0, \xi^\dagger_R \,(\tilde{R} + d) \,\xi_R, 
   \xi^\dagger_R \,d \,\xi_R) \ + \ 
   \rho_4 \,(0, \xi^\dagger_L \,d \,\xi_L,
   \xi^\dagger_L \,(\tilde{L} + d) \,\xi_L ) \nonumber \\
   &\,& \hspace{14mm} + \ \ 
   \rho_4 \,(0, \xi_M \,d \,\xi^\dagger_M, \tilde{L}) \ - \ 
   \rho_4 \,(0, \xi^\dagger_M \, (\tilde{L} + d) \, \xi_M,
   \tilde{R}) \,] \,\, .
\end{eqnarray}
This is a desired effective action, which is expressed in terms of the
dynamical variables $\xi_L$, $\xi_R$, $\xi_M$, $\tilde{L}$ and $\tilde{R}$
of the enlarged hidden local symmetry scheme. (We recall the explicit
form of $\rho_4 (0, A_L, A_R)$ given in (2.96).)
The symmetries of the above action are characterized by its responses
to arbitrary gauge variations belonging to ${[U(n)_L \times U(n)]}^{(ext)}$
as well as ${[U(n)_L \times U(n)_R]}^{(HLS)}$. Under the
independent left ($\theta_L \neq 0, \theta_R = 0$) and right
($\theta_L = 0, \theta_R \neq 0$) gauge variations belonging to
${[U(n)_L \times U(n)_R]}^{(ext)}$, we respectively obtain
\begin{eqnarray}
   \delta^{(ext)}_L \,\,\Gamma^{(a)} \ &=& \ \ \ \,\,\,\,c' 
   \,\,\,\int_{S^4} \,\,\,
   \mbox{tr} \,\,d \,\theta_L \,\,\,\frac{1}{2} \,\,( \,d \,L \,\,L \ + \ 
   L \,d \,L \ + \ L^3 \,) \,\, , \\
   \delta^{(ext)}_R \,\,\Gamma^{(a)} \ &=& \ - \,\,\,c' 
   \,\,\,\int_{S^4} \,\,\,
   \mbox{tr} \,\,d \,\theta_R \,\,\,\frac{1}{2} \,\,( \,d \,R \,\,R \ + \ 
   R \,d \,R \ + \ R^3 \,) \,\, .
\end{eqnarray}
This just corresponds to the familiar left-right symmetric form of
anomaly [36-39].
Specializing to global chiral transformations, which dictates that
$d \,\theta_L = d \,\theta_R = 0$, the above variations identically vanish.
This means that $\Gamma^{(a)}$ maintains global chiral symmetry.
On the other hand, one can verify that, under the transformation
belonging to ${[U(n)_L \times U(n)_R]}^{(HLS)}$, $\Gamma^{(a)}$ is
completely invariant, i.e.
\begin{equation}
   \delta^{(HLS)}_L \,\,\Gamma^{(a)} \ \ = \ \ 
   \delta^{(HLS)}_R \,\,\Gamma^{(a)} \ \ = \ \ 0 \,\, .
\end{equation}
Although these symmetry properties are obvious from the construction
explained above, an explicit proof is given in Appendix A, for
completeness, by calculating the gauge variation of each term of
$\Gamma^{(a)}$. At any rate, the effective action (2.104) or (2.105) has
complete gauge invariance under
${[U(n)_L \times U(n)_R]}^{(HLS)}$, so that one can work in
any gauge one wants. Here, an especially interesting
gauge is such that $\xi_L = \xi_M = 1$, and $\xi_R = U$.
In this special gauge, the action (2.105) reduces to
\begin{eqnarray}
   \Gamma^{(a)} \ \ &\longrightarrow& \ \ c' \,\,\int_{B^5} \,\,
   \omega^0_5 \,(0, U^\dagger \,d \,U) \nonumber \\
   &\,& \!\! - \ \ c' \,\,\int_{S^4} \,\,\,
   [ \,\,\rho_4 \,(0,\,U^\dagger \,d \,U, \,R \,) \ + \ 
   \rho_4 \,(0,\,L, \,U \,(R + d) \,U^\dagger \,) \,\,] \,\, .
\end{eqnarray}
This is nothing but the gauged Wess-Zumino-Witten action
in the left-right symmetric regularization scheme [41,42]. (Naturally,
the role of the electroweak fields in the standard action is played by the
hadronic vector and axial-vector field here.) The effective anomalous
action (2.105) can therefore be thought of as a generalization of the
standard gauged Wess-Zumino-Witten action in that it contains two
extra dynamical fields which work to compensate potentially dangerous
gauge anomaly belonging to ${[U(n)_L \times U(n)_R]}^{(HLS)}$.

\vspace{0mm}
\setcounter{equation}{0}
\section{Effective action with electroweak coupling}

\ \ \ \ \ \ In the previous section, we have derived
from the extended NJL model
an effective meson lagrangian containing not only Goldstone bosons but
also hadronic vector and axial-vector mesons. The obtained effective
action, including the intrinsic-parity nonconserving part, is shown
to have the ${[U(n)_L \times U(n)_R]}^{(HLS)}_{local}$ symmetry as well as
the ${[U(n)_L \times U(n)_R]}^{(ext)}_{global}$ symmetry.
Since the enlarged hidden local symmetry is completely maintained
even at the quantum level, the hadronic vector and axial-vector mesons
in this effective lagrangian can be regarded as gauge bosons of this
symmetry.
On the other hand, the global chiral symmetry possessed by the original
extended NJL lagrangian is maintained by choosing the left-right
symmetric form of anomaly under formally enlarged local symmetry
belonging to ${[U(n)_L \times U(n)_R]}^{(ext)}$.
This choice is physically a natural one, since it respects the global
chiral symmetry, a fundamental symmetry of strong interactions. 
So far, everything goes well without any trouble.
However, once the couplings with
the external electroweak fields is introduced, a nontrivial problem arises.
In this section, we shall explain this problem together with its possible
resolution by paying special attention to the role of the hidden local
symmetry in our effective lagrangian.

The electroweak interactions can be introduced into the
extended NJL model through the standard minimal replacement
in the quark kinetic part of the lagrangian (*) :
\begin{equation}
   \bar{q} \,\,i \,\gamma^\mu \,\,\partial_\mu \,\,q \ \ \longrightarrow
   \ \ \bar{q} \,\,i \,\gamma^\mu \,\,( \,\partial_\mu \ + \ 
   l_\mu \,\,P_L \ + \ r_\mu \,P_R \,) \,\,q \ \ + \ \ {\cal L}_{gauge}
   \,\, ,
\end{equation}
where ${\cal L}_{gauge}$ represents the lagrangian of $SU(2)_L \times U(1)$
electroweak gauge theory, which contains kinetic terms of the external
gauge fields and Higgs fields etc. (We shall omit ${\cal L}_{gauge}$
in the following expressions.) The external gauge fields $l_\mu$ and
$r_\mu$ are expressed in terms of the photon (${\cal B}_\mu$) and
weak bosons (${\cal W}^{\pm}_\mu$ and ${\cal Z}^0_\mu$) as
\begin{eqnarray}
   l_\mu \ \ &=& \ \ i \,e \,Q \,\,( \,{\cal B}_\mu \ - \ 
   \tan \theta_W \,\,{\cal Z}^0_\mu \,) \nonumber \\
   &\,& + \ i \,\,\frac{e}{\sin \theta_W \,\,\cos \theta_W} \,\,T^3 \,\,
   {\cal Z}^0_\mu \ \ + \ \ i \,\,\frac{e}{\sqrt{2} \,\sin \theta_W}
   \,\,{\cal C} \,\,{\cal W}_\mu \,\, ,\\
   r_\mu \ \ &=& \ \ i \,e \,Q \,\,( \,{\cal B}_\mu \ - \ 
   \tan \theta_W \,\,{\cal Z}^0_\mu \,) \,\, ,
\end{eqnarray}
where $\theta_W$, $T^3$ and $Q$ are respectively the Weinberg angle, the
third component of weak isospin and the electric charge. Finally, $C$ is
the generalized Cabbibo matrix.

We can now proceed along the same line as described in sect.2.
First introduce the collective meson field through the addition of (2.3).
Then, instead of (2.6), we obtain
\begin{eqnarray}
   {\cal L} \, &=& \, \bar{q} \,\,[ \,i \,\gamma^\mu \,\{ \,\partial_\mu
   \ + \ (L_\mu + l_\mu) \,P_L \ + \ (R_\mu + r_\mu) \,P_R \,\} \ - \ 
   (M^\dagger \,P_L \, + \, M \,P_R) \,] \,\,q \nonumber \ \ \\
   &\,& \ - \,\,\,
   \frac{1}{4 \,G_S} \,\,\mbox{tr} \,\,M \,M^\dagger \ - \ 
   \frac{1}{8 \,G_V} \,\,\mbox{tr} \,\,[ \,L^2_\mu \ + \ R^2_\mu \,] \,\,.
\end{eqnarray}
Again redefining the quark fields through the chiral rotation (2.23), we
obtain
\begin{eqnarray}
   &\,& \!\!\! \bar{q} \,\,i \,\gamma^\mu \,\,\{ \,\partial_\mu \ + \ 
   (L_\mu + l_\mu) \,P_L \ + \ (R_\mu + r_\mu) \,P_R \,\} \,\,q
   \nonumber \\
   &=& \ \ {\bar{\chi}}_L \,\,i \,\gamma^\mu \,\{ \,\partial_\mu \ + \ 
   \,\xi_L \,\,( \,\partial_\mu + L_\mu + l_\mu \,) \,\xi^\dagger_L \,\}
   \,\,\chi_L \nonumber \\
   &=& \ \ {\bar{\chi}}_R \,\,i \,\gamma^\mu \,\{ \,\partial_\mu \ + \ 
   \xi_R \,\,( \,\partial_\mu + R_\mu + r_\mu \,) \,\xi^\dagger_R \,\}
   \,\,\chi_R \,\, .
\end{eqnarray}
Since $L_\mu$ and $l_\mu$ ($R_\mu$ and $r_\mu$) appear here in the form
$L_\mu + l_\mu$ ($R_\mu + r_\mu$), we find it convenient to introduce the
following redefinition :
\begin{eqnarray}
   {\cal L}_\mu \ \ &=& \ \ L_\mu \ \ + \ \ l_\mu \,\, , \ \ \\
   {\cal R}_\mu \ \ &=& \ \ R_\mu \ \ + \ \ r_\mu \,\, .
\end{eqnarray}
We also introduce chirally rotated fields of ${\cal L}_\mu$ and
${\cal R}_\mu$ by
\begin{eqnarray}
   \tilde{\cal L}_\mu \ \ &=& \ \ \xi_L \,\,( \,{\cal L}_\mu \ + \ 
   \partial_\mu \,) \,\,\xi^\dagger_L \,\, ,\\
   \tilde{\cal R}_\mu \ \ &=& \ \ \xi_R \,\,( \,{\cal R}_\mu \ + \ 
   \partial_\mu \,) \,\,\xi^\dagger_R \,\, ,
\end{eqnarray}
Now using these new variables together with the previously introduced
representations (2.8) and (2.15) with $H \,\simeq \,\,<H> \,\,= \,m$,
(3.4) can be recast into the form :
\begin{eqnarray}
   {\cal L} \ &=& \ \chi \,\,[ \,i \,\gamma^\mu \,( \,\partial_\mu
   \, + \, \tilde{{\cal L}}_\mu \,P_L \, + \,
   \tilde{{\cal R}}_\mu \,P_R \,) \ - \ m \,\,
   ( \, \xi^\dagger_M \,P_L \ + \ \xi_M \, P_R \,) \,] \,\chi \nonumber \\
   &\,& \ - \,\,\,\frac{1}{8 \,G_V} \,\,\mbox{tr} \,\,{[\,\,
   {\tilde{\cal L}}_\mu \ \, + \ \,
   \partial_\mu \,\xi_L \,\xi^\dagger_L \ - \ 
   \xi_L \,l_\mu \,\xi^\dagger_L \,]}^2 \nonumber \\
   &\,& \ - \,\,\,\frac{1}{8 \,G_V} \,\,\mbox{tr} \,\,{[\,\,
   {\tilde{\cal R}}_\mu \ + \ \partial_\mu \,\xi_R \,\xi^\dagger_R \ - \ 
   \xi_R \,r_\mu \,\xi^\dagger_R \,]}^2 \,\, .
\end{eqnarray}
After performing an approximate bosonisation procedure just as before,
we are then led to the following effective lagrangian for the
nonanomalous part :
\begin{eqnarray}
   {\cal L}^{(n)} \ &=& \ \frac{1}{4 \,g^2_V} \,\,\mbox{tr} \,\,
   [ \,\,{\tilde{{\cal L}}}^2_{\mu \nu} \ + \ 
   {\tilde{{\cal R}}}^2_{\mu \nu} \,\,] \ - \ 
   \frac{a}{a - 1} \,\,\frac{f^2_\pi}{4} \,\,\mbox{tr} \,\,
   {[ \,D_\mu \,\xi_M \,\cdot \,\xi^\dagger_M \,]}^2 \nonumber \\
   &\,& \!\!\! - \ \ \frac{1}{2} \,\,a \,\,f^2_\pi \,\,\mbox{tr} \,\,
   {[ \,D_\mu \,\xi_L \,\cdot \,\xi^\dagger_L \,]}^2 \ \ - \ \ 
   \frac{1}{2} \,\,a \,\,f^2_\pi \,\,\mbox{tr} \,\,
   {[ \,D_\mu \,\xi_R \,\cdot \,\xi^\dagger_R \,]}^2 \,\, .
\end{eqnarray}
where
\begin{eqnarray}
   {\tilde{{\cal L}}}_{\mu \nu} \ &=& \ \partial_\mu \,\,
   {\tilde{{\cal L}}}_\nu
   \ \,- \ \,\partial_\nu \,\,{\tilde{{\cal L}}}_\mu \ \,+ \ \,
   [ \,{\tilde{{\cal L}}}_\mu , \,{\tilde{{\cal L}}}_\nu \,]
   \,\, , \\
   {\tilde{{\cal R}}}_{\mu \nu} \ &=& \ \partial_\mu \,\,
   {\tilde{{\cal R}}}_\nu
   \ - \ \partial_\nu \,\,{\tilde{{\cal R}}}_\mu \ \,+ \ \,
   [ \,{\tilde{{\cal R}}}_\mu , \,{\tilde{{\cal R}}}_\nu \,] 
   \,\, , \ 
\end{eqnarray}
and
\begin{eqnarray}
   D_\mu \,\,\xi_L \ \ &=& \ \ \partial_\mu \,\,\xi_L \ + \ 
   {\tilde{{\cal L}}}_\mu \,\,\xi_L \ - \ \xi_L \,\,l_\mu \,\, , \\
   D_\mu \,\,\xi_R \ \ &=& \ \ \partial_\mu \,\,\xi_R \ + \ 
   {\tilde{{\cal R}}}_\mu \,\,\xi_R \ - \ \xi_R \,\,r_\mu \,\, , \\
   D_\mu \,\,\xi_M \ \ &=& \ \ \partial_\mu \,\,\xi_M \ + \ 
   {\tilde{{\cal L}}}_\mu \,\,\xi_M \ - \ \xi_M \,\,
   {\tilde{{\cal R}}}_\mu \,\, .
\end{eqnarray}
Identifying ${\tilde{\cal L}}_\mu$ and ${\tilde{\cal R}}_\mu$ as the
gauge bosons of the enlarged hidden local symmetry, the above lagrangian
precisely coincides with the corresponding lagrangian of Bando et al.
with the electroweak couplings except that the arbitrary constants
$a'$, $b'$, $c'$, $d'$ of their model are again constrained such that
$a' = b' = a$, $c' = a / (a-1)$ and $d' = 0$ in our effective lagrangian.
As already discussed by Bando et al. [6], an especially interesting
gauge corresponds to taking $\xi_L = \xi_R = 1$, $\xi_M = U$.
In this particular gauge, (3.11) reduces to
\begin{eqnarray}
   {\cal L}^{(n)} \ &=& \ \frac{1}{4 \,g^2_V} \,\,\mbox{tr} \,\,
   [\,{{\cal L}}^2_{\mu \nu} \ + \ {{\cal R}}^2_{\mu \nu} \,]
   \ + \ \frac{a}{a-1} \,\,\frac{f^2_\pi}{4} \,\,\mbox{tr} \,\,
   ( \,D_\mu \,\,U \,\,D^\mu \,\,U^\dagger \,) \nonumber \\
   &\,& \!\!\! - \ \ \frac{1}{2} \,\,a \,\,f^2_\pi \,\,\mbox{tr} \,\,
   {( \,{{\cal L}}_\mu - l_\mu \,)}^2 \ - \ 
   \frac{1}{2} \,\,a \,\,f^2_\pi \,\,\mbox{tr} \,\,
   {( \,{{\cal R}}_\mu - r_\mu \,)}^2 \,\, , \ \ \
\end{eqnarray}
with
\begin{eqnarray}
   D_\mu \,\,U \ \ &\equiv& \ \ \partial_\mu \,\,U \,\,\ + \ 
   {\cal L}_\mu \,\,U \,\,\ - \ U \,\,{{\cal R}}_\mu \,\, ,\\
   D_\mu \,\,U^\dagger \ \,\, &\equiv& \ \ 
   \partial_\mu \,\,U^\dagger \ - \ 
   U^\dagger \,\,{\cal L}_\mu \ + \ 
   {\cal R}_\mu \,\, U^\dagger \,\, .
\end{eqnarray}
Identifying ${{\cal L}}_\mu$ and ${{\cal R}}_\mu$ as physical
vector and axial-vector mesons, (3.17) is essentially the lagrangian of
the massive Yang-Mills model supplemented by the VMD-type direct
couplings between the hadronic vector and axial-vector mesons and the
external gauge fields $l_\mu$ and $r_\mu$ [1,2].

Now we turn to more interesting anomalous part of the action.
For a pedagogical reason, we first show how the naive quantization
procedure cause a trouble, and then explain how the trouble can be
circumvented following the recent proposal by Bijinens and Prades and by
Arriola and Salcedo.
As is clear from the discussion in sect.2, the fermion path integral
$Z_f$ can be written in the following forms :
\begin{eqnarray}
   Z_f \ \ &=& \ \ \int \,\,{\cal D} q \,\,{\cal D} \bar{q} \,\,\,\,
   e^{\,i \,\int \,\,d^4 x \,\,\bar{q} \,\,{\cal D} \,\,q} \nonumber \\
   &=& \ \ {J_1}^{N_c} \,\,\int \,\,
   {\cal D} \chi \,\,{\cal D} \bar{\chi} \,\,\,\,
   e^{\,i \,\int \,\,d^4 x \,\,\bar{\chi} \,\,\tilde{{\cal D}} \,\,\chi}
   \nonumber \\
   &=& \ \ {J_1}^{N_c} \,\,{J_2}^{N_c} \,\,\int \,\,
   {\cal D} \varphi \,\,{\cal D} \bar{\varphi} \,\,\,\,
   e^{\,i \,\int \,\,d^4 x \,\,\bar{\varphi} \,\,
   \hat{{\cal D}} \,\,\varphi} \,\, ,
\end{eqnarray}
where
\begin{eqnarray}
   {\cal D} \, &=& \, i \,\gamma^\mu \,
   ( \,\partial_\mu \, + \, {\cal L}_\mu \,\,P_L \, + \,
   {\cal R}_\mu \,\,P_R \,) \ - \ 
   m \,( \,\xi^\dagger_R \,\,\xi^\dagger_M \,\,\xi_L \,\,P_L \, + \,
   \xi^\dagger_L \,\,\xi_M \,\,\xi_R \,\,P_R \,) \,\, ,\ \ \\
   \tilde{{\cal D}} \, &=& \, i \,\gamma^\mu \,
   ( \,\partial_\mu \, + \, \tilde{{\cal L}}_\mu \,\,P_L \, + \,
   \tilde{{\cal R}}_\mu \,\,P_R \,) \ - \ 
   m \,( \,\xi^\dagger_M \,\,P_L \, + \,
   \xi_M \,\,P_R \,) \,\, ,\\
   \hat{{\cal D}} \, &=& \, i \,\gamma^\mu \,
   ( \,\partial_\mu \, + \, \hat{{\cal L}}_\mu \,\,P_L \, + \,
   \hat{{\cal R}}_\mu \,\,P_R \,) \ - \ m \,\, .
\end{eqnarray}
which have the same forms as (2.37) $\sim$ (2.40) except that $L_\mu$ and
$R_\mu$ and their chirally rotated fields $\tilde{L}_\mu$,
$\tilde{R}_\mu$, $\hat{L}_\mu$, $\hat{R}_\mu$ there
are now replaced by ${{\cal L}}_\mu \equiv L_\mu + l_\mu$,
${{\cal R}}_\mu \equiv R_\mu + r_\mu$ and their chirally rotated
correspondents. It is therefore quite natural to think that the
anomalous action with the external electroweak couplings is obtained from
(2.60) in the previous section simply by replacing
$L_\mu$, $R_\mu$, $\tilde{L}_\mu$, $\tilde{R}_\mu$, $\hat{L}_\mu$,
$\hat{R}_\mu$ by ${\cal L}_\mu$, ${\cal R}_\mu$,
$\tilde{{\cal L}}_\mu$, $\tilde{{\cal R}}_\mu$, $\hat{{\cal L}}_\mu$,
$\hat{\cal {R}}_\mu$. However, it turns out that this simplest
construction does not meets the requirement of QCD phenomenology.
To see it, we first recall the symmetries possessed by such an action.
Its symmetries are ${[U(n)_L \times U(n)_R]}^{(ext)}_{global} \times
{[U(n)_L \times U(n)_R]}^{(HLS)}$. Remember that the global chiral
symmetry here is the consequence of our choice of the left-right
symmetric regularization scheme.
In the absence of the electroweak couplings, this choice has nothing to be
questioned, since it respects the global chiral symmetry, i.e.
the fundamental symmetry of strong interactions.
However, since the electromagnetic gauge group is now contained in the
diagonal subgroup ${[U(n)_V]}^{(ext)}$ of ${[U(n)_L \times
U(n)_R]}^{(ext)}$, the anomalous action in the left-right symmetric
regularization scheme breaks electromagnetic gauge invariance.
This is nothing but the problem several authors had
encountered when trying to construct the anomalous action based on
the idea of gauging the external chiral symmetry [28-31].
To recover the electromagnetic gauge invariance, they then decided to
adopt the vector-gauge invariant regularization scheme, which is
attained by subtracting a local counter term (called the Bardeen
subtraction) depending on the hadronic vector and axial-vector
fields. This cannot get rid of the trouble, however.
The gauged Wess-Zumino-Witten action in the vector-gauge invariant scheme
inevitably breaks the global chiral symmetry at the strong
interaction level in the present setting.
Consequently, the famous
low-energy theorem for the purely hadronic process $K^{+} \, K^{-}
\longrightarrow 3 \,\pi$ is not correctly reproduced. It was also
shown [22] that the above symmetry violation,
in combination with the mixing
of the Goldstone boson and the axial-vector field, brings about
theoretically unpleasant
correction to the process $\gamma \longrightarrow 3 \,\pi$.
It seemed that there is no way out of this dilemma, considering that
the anomaly can be shifted from one place to another but it cannot be
eliminate completely.

However, here is a pitfall. It was an implicit assumption of the argument
so far that the local counter terms in this anomaly shifting procedure
are function of ${\cal L}_\mu$ and ${\cal R}_\mu$ only.
As has been pointed out by Bijinens and Prades quite recently, this
may not be necessarily true [32]. According to them, this corresponds to
the standardly used procedure in which the functional measure of the
hadronic vector and axial-vector fields is defined by the Dirac operator
(3.21) that is a function of ${\cal L}_\mu$ and ${\cal R}_\mu$
rather than a function of $L_\mu$ and $R_\mu$, which has no {\it a priori}
justification. This observation opens up a possibility to use more
general renormalization procedure. That is one is now allowed to
subtract local counter terms, which has general dependence on $L_\mu$ and
$l_\mu$ (and $R_\mu$ and $r_\mu$), in the
construction of the effective meson action. Recently, Arriola and Salcedo
has made use of this observation for explicitly constructing the anomalous
action with the required symmetries [33].
Here we shall carry out a similar construction in our scheme with
the extra hidden local symmetry.

We start with the anomalous action :
\begin{eqnarray}
   \Gamma^{(a)}_{LR} \ 
   &=& \ c' \,\,\,\int_{B^5} \,\,\,[ \,\,
   \omega^0_5 \,(\, 0, \,\xi^\dagger_R \,d \xi_R) \ - \ 
   \omega^0_5 \,(\, 0, \,\xi^\dagger_L \,d \xi_L) \ - \ 
   \omega^0_5 \,(\, 0, \,\xi_M \,d \,\xi^\dagger_M) \,\,] \nonumber \\
   &+& \ c' \,\,\,\int_{S^4} \,\,\,[ \,\,
   \rho_4 \,(\,0, \,{\cal R}, \,\xi^\dagger_R \,d \,\xi_R) \ + \ 
   \rho_4 \,(\,0, \,\xi^\dagger_L \,d \,\xi_L, \,{\cal L}) \nonumber \\
   &\,& \hspace{15mm} + \ \ 
   \rho_4 \,(\,0, \,\xi_M \,d \,\xi^\dagger_M, \,\tilde{{\cal L}}) \ - \ 
   \rho_4 \,(\,0, \,\hat{{\cal L}}, \,\hat{\cal{R}}) \,] \,\, ,
\end{eqnarray}
which is obtained from (2.104) simply by replacing 
$L_\mu$, $R_\mu$, $\tilde{L}_\mu$, $\tilde{R}_\mu$, $\hat{L}_\mu$,
$\hat{R}_\mu$ by ${\cal L}_\mu$, ${\cal R}_\mu$,
$\tilde{{\cal L}}_\mu$, $\tilde{{\cal R}}_\mu$, $\hat{{\cal L}}_\mu$,
$\hat{\cal {R}}_\mu$. As will be shown in Appendix A, this effective
action has the following anomaly structure :
\begin{eqnarray}
    \delta^{(ext)}_L \,\,\Gamma^{(a)}_{LR} \ \ &=& \ \ 
    G_L \,\,(\,\theta_L \,; \,{\cal L}, \,{\cal R}) \ \ = \ \ \ 
    \,\,\,\,\,c' \,\,\,\int_{S^4} \,\,\,
    \bar{\Delta} \,(\,\theta_L, {\cal L} \,) \,\, ,\\
    \delta^{(ext)}_R \,\,\Gamma^{(a)}_{LR} \ \ &=& \ \ 
    G_R \,\,(\,\theta_R \,; \,{\cal L}, \,{\cal R}) \ \ = \ \ 
    - \,\,c' \,\,\,\int_{S^4} \,\,\,
    \bar{\Delta} \,(\,\theta_R, {\cal R} \,) \,\, ,\\
    \delta^{(HLS)}_L \,\,\Gamma^{(a)}_{LR} \ &=& \ \ 
    \delta^{(HLS)}_R \,\,\Gamma^{(a)}_{LR} \ \ = \ \ 0 \,\, .
\end{eqnarray}
We shall see below that allowing general counter terms, which depends on
${\cal L}_\mu$ and $r_\mu$ (and ${\cal R}_\mu$ and
$l_\mu$), one can shift the anomaly to the external electroweak sector.
Before doing this, we should add two remarks, which would help to avoid
confusion. First, it is important to recognize the fact that in the
effective action $\Gamma^{(a)}_{LR}$ the anomaly resides only in the
${[U(n)_L \times U(n)_R]}^{(ext)}$ group, and the
$[{U(n)_L \times U(n)_R]}^{(HLS)}$ group is anomaly free. The subtraction
of local counter terms, which depends on ${\cal L}_\mu$ and
${\cal R}_\mu$, does not change the anomaly-free nature of the
$[{U(n)_L \times U(n)_R]}^{(HLS)}$ group, since ${\cal L}_\mu$ and
${\cal R}_\mu$ are absolutely intact under this group transformations.
Secondly, as counter terms, we choose a function of ${\cal L}_\mu$
and $l_\mu$ (and ${\cal R}_\mu$ and $r_\mu$) instead of the choice
in [33], where it is chosen as a function of $L_\mu$ and $l_\mu$
(and $R_\mu$ and $r_\mu$), where ${\cal L}_\mu = L_\mu + l_\mu$
(and ${\cal R}_\mu = R_\mu + r_\mu$).
This however makes no essential difference, since a function of
$L_\mu$ and $l_\mu$ can trivially be expressed as (another) function
of ${{\cal L}}_\mu = L_\mu + l_\mu$ and $l_\mu$.
Our choice here is motivated by the fact that not $L_\mu$ and $R_\mu$ but
${\cal L}_\mu$ and ${\cal R}_\mu$
(or ${\tilde{\cal L}}_\mu$ and ${\tilde{\cal R}}_\mu$)
should be identified with
the physical fields in order to eliminate the kinetic term mixing
resulting from the bosonisation of (3.4).

Now we define the new action from (3.24) by subtracting an appropriate
local counter term as
\begin{eqnarray}
   {\bar{\Gamma}}^{(a)} \ \ \equiv \ \ \Gamma^{(a)}_{LR} \ \ - \ \ 
   \Gamma^{(a)}_{c.t.} \,\, .
\end{eqnarray}
Here we require that the counter term is globally chiral invariant, so that
the subtraction of it preserves this symmetry.
There are eight globally chiral invariant pieces with dimension four,
which can be constructed from ${\cal L}$ and $l$. They are
\begin{eqnarray}
   &\,& \mbox{tr} \,\,[\,{\cal L}^3 \,l \,] \, , \ \ \ \ \ 
   \mbox{tr} \,\,[\,{\cal L}^2 \,\,l^2 \,] \, , \ \ \ \ \ 
   \mbox{tr} \,\,[\,{\cal L} \,\,l \,\,
   {\cal L} \,\,l \,] \, , \ \ \ \ \ 
   \mbox{tr} \,\,[\,{\cal L} \,\,l^3 \,] \, , \nonumber \\ 
   &\,& \mbox{tr} \,\,[\,d \,{\cal L} \,\,{\cal L} \,\,
   l \,] \, , \ \ 
   \mbox{tr} \,\,[\,{\cal L} \,\,
   d \,{\cal L} \,\,l \,] \, , \ \ 
   \mbox{tr} \,\,[\,{\cal L} \,\, d \,l \,\,l \,] \, , \ \ 
   \mbox{tr} \,\,[\,{\cal L} \,\,l \,\, d \,l \,] \,\, .
\end{eqnarray}
We then rewrite $\Gamma^{(a)}_{c.t.}$ as a linear combination of these
quantities and their right-handed counterparts as
\begin{eqnarray}
   \Gamma^{(a)}_{c.t.} \ &=& \ - \,\,\frac{c'}{2} \,\,\,\int_{S^4} \,\,\,
   \mbox{tr} \,\,[ \,\,
   c_1 \,\,{\cal L} \,\,l^3 \ + \ 
   c_2 \,\,{\cal L} \,\, d \,l \,\,l \ + \ 
   c_3 \,\,{\cal L} \,\,l \,\, d \,l \ + \ 
   c_4 \,\,{\cal L}^2 \,\,l^2 \nonumber \\
   &\,& \hspace{15mm} \ + \ \ 
   c_5 \,\,{\cal L} \,\,l \,\,{\cal L} \,\,l \ + \ 
   c_6 \,\,d \,{\cal L} \,\,{\cal L} \,\,l \ + \ 
   c_7 \,\,{\cal L} \,\, d \,{\cal L} \,\,l \ + \ 
   c_8 \,\,{\cal L}^3 \,\,l \,] \nonumber \\
   &\,& \ \ - \ \ \ ( \,{\cal L} \leftrightarrow {\cal R}, \ 
   l \leftrightarrow r \,) \,\, .
\end{eqnarray}
The eight unknown coefficients $c_1, \cdots ,c_8$ can be determined such
that the new action $\bar{\Gamma}^{(a)}$ satisfies the following
conditions :
\begin{eqnarray}
   \delta^{(ext)}_L \,\,{\bar{\Gamma}}^{(a)} \ \ &=& \ \ 
   G_L \,( \,\theta_L \ ; \ l, \,r \,) \,\, ,\\
   \delta^{(ext)}_R \,\,{\bar{\Gamma}}^{(a)} \ \ &=& \ \ 
   G_R \,( \,\theta_L \ ; \ l, \ r \,) \,\, ,\\
   \delta^{(HLS)}_L \,\,{\bar{\Gamma}}^{(a)} \ \ &=& \ \ 
   \delta^{(HLS)}_R \,\,{\bar{\Gamma}}^{(a)} \ \ = \ \ 0 \,\, .
\end{eqnarray}
Here, the third condition is trivially satisfied, since
the subtracted counter term is hidden gauge invariant.
The first and the second conditions can alternatively be expressed as
\begin{eqnarray}
   \delta^{(ext)}_L \,\,\Gamma^{(a)}_{c.t.} \ \ &=& \ \ 
   G_L \,(\,\theta_L \,; \,{\cal L}, \,{\cal R} \,) \ - \ 
   G_L \,(\,\theta_L \,; \,l, \,r \,) \,\, ,\\
   \delta^{(ext)}_R \,\,\Gamma^{(a)}_{c.t.} \ \ &=& \ \ 
   G_R \,(\,\theta_R \,; \,{\cal L}, \,{\cal R} \,) \ - \ 
   G_R \,(\,\theta_R \,; \,l, \,r \,) \,\, ,
\end{eqnarray}
which is fulfilled if one take as
\begin{equation}
   c_1 \ = \ c_2 \ = \ c_3 \ = \ 1, \ \ c_4 \ = \ 0, \ \ 
   c_5 \ = \ 1 / 2, \ \ c_6 \ = \ c_7 \ = \ c_8 \ = \ 1 \,\, ,
\end{equation}
which gives the desired counter term :
\begin{eqnarray}
   \Gamma^{(a)}_{c.t.} \ &=& \ - \,\,\frac{c'}{2}
   \,\,\,\int_{S^4} \,\,\mbox{tr} \,\,
   [ \,\,{\cal L} \,\,l^3 \ + \ {\cal L} \,\,\{ \,dl, l \,\}
   \ + \ \frac{1}{2} \,\,{\cal L} \,\,l \,\,{\cal L} \,\,l
   \nonumber \\
   &\,& \hspace{17mm} + \ \ 
   \{ \,d \,{\cal L}, {\cal L} \,\} \,\,l
   \ + \ {\cal L}^3 \,\,l \,\,] \ \ - \ \ 
   ( \,{\cal L} \leftrightarrow {\cal R}, \ 
   l \leftrightarrow r \,) \,\, .
\end{eqnarray}
Now one sees from (3.31) and (3.32) that the anomaly is totally shifted
from the hadronic sector to the external electroweak sector.
To recover the electromagnetic gauge invariance completely, we need further
redefinition of the anomalous action as
\begin{eqnarray}
   \Gamma^{(a)} \ \ \ \equiv \ \ \ {\bar{\Gamma}}^{(a)} \ \ - \ \ 
   \Gamma^{(a)}_{LR} \,\,[ \,\,\xi_L \,= \,\xi_R \,= \,\xi_M \,= \,1
   \, ; \, l, \,r \,] \,\, ,
\end{eqnarray}
where the subtracted term above is nothing but the familiar Bardeen
subtraction, which depends on the external gauge fields $l$ and $r$
instead of the hadronic fields ${\cal L}$ and ${\cal R}$.
Now, the anomaly structure of our final action is given by
\begin{eqnarray}
   \delta^{(ext)}_V \,\,\Gamma^{(a)} \ \ &=& \ \ 0 \,\, ,\\
   \delta^{(ext)}_A \,\,\Gamma^{(a)} \ \ &=& \ \ 
   G_B \,( \,\theta_R = - \,\,\theta_L = \theta \, ; \,v, \,a \,) \,\, ,\\
   \delta^{(HLS)}_V \,\,\Gamma^{(a)} \ &=& \ \ 
   \delta^{(HLS)}_A \,\,\Gamma^{(a)} \ \ = \ \ 0 \,\, .
\end{eqnarray}
As a matter of course, there still is an axial anomaly, which depends
on the external electroweak fields. However, it is
the standardly accepted scenario that it is to be canceled by the
corresponding lepton loop contribution owing to the
quark-lepton symmetry. Now our final action satisfies all the symmetries
as required as an effective theory of QCD (except for $U_A (1)$ anomaly).
For convenience, we summarize the final form of the effective action
derived here from the extended NJL model with inclusion of the enlarged
hidden gauge symmetry :
\begin{equation}
   \Gamma_{eff} \ \ = \ \ \Gamma^{(n)} \ \ + \ \ \Gamma^{(a)} \,\, ,
\end{equation}
where $\Gamma^{(n)} \, = \, \int \,\,d^4 x \,\,{\cal L}^{(n)}$ with
\begin{eqnarray}
   {\cal L}^{(n)} \ \ &=& \ \ \frac{1}{4 \,g^2_V} \,\,\mbox{tr} \,\,
   [ \,{\tilde{{\cal L}}}^2_{\mu \nu} \ \ + \ \ 
   {\tilde{{\cal R}}}^2_{\mu \nu} \,]
   \ \ - \ \ \frac{a}{a - 1} \,\,\frac{f^2_\pi}{4} \,\,\mbox{tr} \,\,
   {[ \,D_\mu \,\xi_M \,\cdot \,\xi^\dagger_M \,]}^2 \nonumber \\
   &\,& - \ \ \frac{1}{2} \,\,a \,\,f^2_\pi \,\,\mbox{tr} \,\,
   {[ \,D_\mu \,\xi_L \,\cdot \,\xi^\dagger_L \,]}^2 \ \ - \ \ 
   \frac{1}{2} \,\,a \,\,f^2_\pi \,\,\mbox{tr} \,\,
   {[ \,D_\mu \,\xi_R \,\cdot \,\xi^\dagger_R \,]}^2 \,\, .
\end{eqnarray}
and
\begin{eqnarray}
   \Gamma^{(a)} \ \ &=& \ \ \Gamma^{(a)}_{LR} \,
   [\,\xi_L,\,\xi_R,\,\xi_M \,\, ; \,
   \tilde{\cal L}, \tilde{\cal R} \,] \nonumber \\
   &\,& - \ \ \Gamma^{(a)}_{c.t.} \,
   [ \,{\cal L},{\cal R} \, ; \, l, r \,] \ \ - \ \ 
   \Gamma^{(a)}_{LR} \,[\,\xi_L = \xi_R = \xi_M = 1 \,; \, l,\, r \,] \,\, .
\end{eqnarray}
Taking the special gauge $\xi_L = \xi_M = 1$, $\xi_R = U$,
$\Gamma^{(a)}_{LR} \,[\,\xi_L, \xi_R, \xi_M \,; \,\tilde{L}, \tilde{R} \,]$
reduces to the standard gauged Wess-Zumino-Witten action in the
LR scheme, i.e. $\Gamma^{(WZ)}_{LR} \,[\,U \,;\,\cal{L}, \cal{R} \,]$
given in terms of the field variables $U$, ${\cal L}$ and ${\cal R}$,
while $\Gamma^{(a)}_{LR} \,[\,\xi_L = \xi_R = \xi_M = 1 \,; \,l,r \,]$ is
nothing different from $\Gamma^{(WZ)}_{LR} \,[\,U=1 \,; \,l, r \,]$.
The total action then becomes
\begin{eqnarray}
   {\cal L}^{(n)} \ \ &=& \ \ \frac{1}{4 \,g^2_V} \,\,\mbox \,\,
   [\,{\cal L}^2_{\mu \nu} \ + \ {\cal R}^2_{\mu \nu} \,]
   \ + \ \frac{a}{a-1} \,\,\frac{f^2_\pi}{4} \,\,\mbox{tr} \,\,
   ( \,D_\mu \,\,U \,\,D^\mu \,\,U^\dagger \,) \,\, \nonumber \\
   &\,& - \ \ \frac{1}{2} \,\,a \,\,f^2_\pi \,\,\mbox{tr} \,\,
   {( \,{\cal L}_\mu - l_\mu \,)}^2 \ - \ 
   \frac{1}{2} \,\,a \,\,f^2_\pi \,\,\mbox{tr} \,\,
   {( \,{\cal R}_\mu - r_\mu \,)}^2 \,\, ,
\end{eqnarray}
and
\begin{equation}
   \Gamma^{(a)} \ \ = \ \ \Gamma^{(WZ)}_{LR} \,
   [ \,U \,; \,{\cal L}, \,{\cal R} \,] \ - \ 
   \Gamma^{(a)}_{c.t.} \,[ \,{\cal L}, {\cal R} \,; \,l, r \,] \ - \ 
   \Gamma^{(WZ)}_{LR} \,[\,U=1 \,; \,l, r \,] \,\, ,
\end{equation}
which essentially coincides with the result of Arriola and Salcedo
except that we express the action in terms of ${\cal L}$ and $l$
(${\cal R}$ and $r$), while they do in terms of $L = {\cal L} - l$
and $l$ ($R = {\cal R} - r$ and $r$).

It seems clear by now that the idea of the hidden local symmetry plays no
positive role in the above construction of the phenomelologically
consistent effective action. What is important from the physical
viewpoint is response of the action
under the external global or local variations. This seems reasonable
because the structure of hadronic currents or their associated observables
can be seen only with the external electroweak probes.
In the following discussion on the theoretical structure of the hadronic
currents, we therefore concentrate on the hidden gauge fixed version of
the effective action, for convenience.

As first pointed out in [32], the aforementioned subtraction of the
local counter term, which depends on ${\cal L}_\mu$ and $l_\mu$
(and ${\cal R}_\mu$ and $r_\mu$), modifies the vector meson
dominance (VMD), which is otherwise exact in the extended NJL model.
How it is modified can be seen as follows.
To this end, we first divide the total effective action
$\Gamma_{eff} = \Gamma^{(n)} + \Gamma^{(n)}$ into purely hadronic part
and the other part that consists of terms containing at least one external
gauge fields :
\begin{equation}
   \Gamma_{eff} \ \ = \ \ \Gamma_{strong} \ \ + \ \ 
   \mbox{terms containing electroweak fields} \,\, ,
\end{equation}
where $\Gamma_{strong}$ is given by
\begin{equation}
   \Gamma_{strong} \ \ = \ \ \Gamma_f \ - \ \frac{M^2_V}{2 \,g^2_V} \,\,
   \mbox{tr} \,\,[\,\,{\cal L}^2_\mu \, + \,
   {\cal R}^2_\mu \,\,]
   \,\, ,
\end{equation}
with $\Gamma_f$ being the part, which comes from the path integral of
the fermion determinant with appropriate counter term subtraction, i.e.
\begin{equation}
   \Gamma_f \ \ \equiv \ \ - \,i \,\,N_c \,\,\log \,\,\mbox{det} \,\,
   {D (\,U,\,{\cal L},\,{\cal R} \,)\,\,|}_{renorm} \,\, .
\end{equation}
Our lowest order answer has been
\begin{eqnarray}
   \Gamma_f \ \ &=& \ \ \frac{1}{4 \,g^2_V} \,\,\mbox{tr} \,\,
   [\,{\cal L}^2_{\mu \nu} \ + \ {\cal R}^2_{\mu \nu} \,]
   \ + \ \frac{a}{a-1} \,\,\frac{f^2_\pi}{4} \,\,\mbox{tr} \,\,
   (\,D_\mu \,U \,\,D^\mu \,U^\dagger \,) \nonumber \\
   &\,& + \ \ \Gamma^{(WZ)}_{LR} [\,U ; \,{\cal L}, \,{\cal R} \,] 
   \, - \, \Gamma^{(a)}_{c.t.} \,[ \,{\cal L}, \,{\cal R} \,; \,l,\,r \,]
   \, - \, \Gamma^{(WZ)}_{LR} \,[ \,U = 1 \, ; \,l, \,r \,] \,\, . \ \ 
\end{eqnarray}

Now we consider the change of $\Gamma_{strong}$ under an infinitesimal
left variation, which gives
\begin{eqnarray}
   \delta^{(ext)}_L \,\,\Gamma_{strong} \ &=& \ \int \,\,\mbox{tr} \,\,
   \{ \,\delta^{(ext)}_L \,\,{\cal L}_\mu \,\,
   \frac{\delta \,\Gamma_{strong}}{\delta \,{\cal L}_\mu}  \ + \ 
   \delta^{(ext)}_L \,\,U \,\,
   \frac{\delta \,\Gamma_{strong}}{\delta \,U} \,\,\} \nonumber \\
   &=& \ \ \int \,\,\mbox{tr} \,\,
   \{ \,\delta^{(ext)}_L \,\,{\cal L}_\mu \,\,( \,
   \frac{\delta \,\Gamma_f}{\delta \,{\cal L}_\mu} \ - \ 
   \frac{M^2_V}{g^2_V} \,\,{\cal L}_\mu \,) \ + \ 
   \delta^{(ext)}_L \,\,U \,\,
   \frac{\delta \,\Gamma_{strong}}{\delta \,U} \,\,\} \nonumber \\
   &=& \ \ \int \,\,\,\mbox{tr} \,\,\theta_L \,\,
   \{ \,\,- \,\,D^\mu \,({\cal L}) \,\,
   \frac{\delta \,\Gamma_f}{\delta \,{\cal L}_\mu} \ + \ 
   \frac{M^2_V}{g^2_V} \,\,\partial^\mu \,{\cal L}_\mu \ - \ 
   U \,\,\frac{\delta \,\Gamma_{strong}}{\delta \,U} \,\,\} \,\, .
\end{eqnarray}
Here we have performed a partial integration and introduced the covariant
derivative $D^\mu ({\cal L})$ operating on a matrix $M$ by
\begin{equation}
   D^\mu \,\,({\cal L}) \,\,M \ \ = \ \ \partial^\mu \,\,M \ + \ 
   [\,{\cal L}^\mu, M \,] \,\,\, .
\end{equation}
Using the equation of motion for the Goldstone field, i.e.
$\delta \,\Gamma_{strong} \,/ \,\delta U \,= \,0$, we then obtain
\begin{equation}
   \delta^{(ext)}_L \,\,\Gamma_{strong} \ \ = \ \ - \,\,
   \int \,\,\mbox{tr} \,\,\,
   \theta_L \,\,\{ \,D^\mu ({\cal L}) \,\,j^L_\mu \ - \ 
   \frac{M^2_V}{g^2_V} \,\,\partial^\mu \,{\cal L}_\mu \,\} \,\, ,
\end{equation}
where
\begin{eqnarray}
   j^L_\mu \ \ \equiv \ \ 
   \frac{\delta}{\delta {\cal L}_\mu}  \,\,\Gamma_f \,\, ,
\end{eqnarray}
is the basic quark left-hand current (or more precisely its bosonic
equivalent). Similarly, the infinitesimal right variation of
$\Gamma_{strong}$ gives
\begin{equation}
   \delta^{(ext)}_R \,\,\Gamma_{strong} \ \ = \ \ - \,\,
   \int \,\,\mbox{tr} \,\,
   \theta_R \,\,\{ \,D^\mu ({\cal R}) \,\,j^R_\mu \ - \ 
   \frac{M^2_V}{g^2_V} \,\,\partial^\mu \,{\cal R}_\mu \,\} \,\, ,
\end{equation}
with
\begin{eqnarray}
   j^R_\mu \ \ \equiv \ \ 
   \frac{\delta}{\delta {\cal R}_\mu} \,\,\Gamma_f \,\, .
\end{eqnarray}
Here, an important observation is as follows.
The equation of motions for the hadronic fields result from the stationary
requirement of $\Gamma_{strong}$ under arbitrary variations of
${\cal L}_\mu$, ${\cal R}_\mu$, and $U$. Since the gauge
variations (3.53) and (3.55) are special cases of such arbitrary
variations, it immediately follows that
\begin{equation}
   \delta^{(ext)}_L \,\,\Gamma_{strong} \ \ = \ \ 
   \delta^{(ext)}_R \,\,\Gamma_{strong} \ \ = 0 \,\, .
\end{equation}
Combining (3.54) and (3.55) and (3.57), we therefore obtain
\begin{eqnarray}
   D^\mu \,({\cal L}) \,\,j^L_\mu \ \ &=& \ \ \frac{M^2_V}{g^2_V} \,\,
   \partial^\mu \,{\cal L}_\mu \,\, ,\\
   D^\mu \,({\cal R}) \,\,j^R_\mu \ \ &=& \ \ \frac{M^2_V}{g^2_V} \,\,
   \partial^\mu \,{\cal R}_\mu \,\, .
\end{eqnarray}
Incidentally, the covariant derivatives of the basic currents give
anomaly (it can be easily verified by carrying out a similar manipulation
as above for $\Gamma_f$ instead of $\Gamma_{strong}$) as
\begin{eqnarray}
   D^\mu \,({\cal L}) \,\,j^L_\mu \ &=& \ \,- \ \,\partial^\mu \,
   \Delta_\mu \,\,({\cal L}_\mu) \,\, ,\\
   D^\mu \,({\cal R}) \,\,j^R_\mu \ &=& \ \ \ \,\,\,\,\partial^\mu \,
   \Delta_\mu \,\,({\cal R}_\mu) \,\, ,
\end{eqnarray}
with
\begin{eqnarray}
   \Delta^\mu ({\cal L}_\mu) \ \ &=& \ \ - \,\,\frac{c'}{2} \,\,
   \varepsilon^{\mu \nu \rho \sigma} \,\,[ \,
   \{\,{\cal L}_\nu,\,\partial_\rho \,{\cal L}_\sigma \,\} \ \,+ \ \,
   {\cal L}_\nu \,\,{\cal L}_\rho \,\,{\cal L}_\sigma \,] \,\, ,\\
   \Delta^\mu ({\cal R}_\mu) \ \ &=& \ \ - \,\,\frac{c'}{2} \,\,
   \varepsilon^{\mu \nu \rho \sigma} \,\,[ \,
   \{\,{\cal R}_\nu,\partial_\rho \,{\cal R}_\sigma \,\} \ + \ 
   {\cal R}_\nu \,\,{\cal R}_\rho \,\,{\cal R}_\sigma \,] \,\, .
\end{eqnarray}
Remember that the total derivative nature of the anomaly results from our
choice of the left-right symmetric form of anomaly.
Combining (3.58),(3.59), and (3.60),(3.61), we then find
\begin{eqnarray}
   \partial^\mu \,\,( \,\frac{M^2_V}{g^2_V} \,\,{\cal L}_\mu \ + \ 
   \Delta_\mu \,\,({\cal L}_\mu) \,) \ \ \,\,&=& 0 \,\, ,\\
   \partial^\mu \,\,( \,\frac{M^2_V}{g^2_V} \,\,{\cal R}_\mu \ - \ 
   \Delta_\mu \,\,({\cal R}_\mu) \,) \ \ &=& 0 \,\, ,
\end{eqnarray}
which insists the existence of the conserved currents given by
\begin{eqnarray}
   {\bar{J}}^L_\mu \ \ &\equiv& \ \ \frac{M^2_V}{g^2_V} \,\,{\cal L}_\mu
   \ \ + \ \ \Delta_\mu \,\,({\cal L}_\mu) \,\, ,\\
   {\bar{J}}^R_\mu \ \ &\equiv& \ \ \frac{M^2_V}{g^2_V} \,\,{\cal R}_\mu
   \ \ - \ \ \Delta_\mu \,\,({\cal R}_\mu) \,\, .
\end{eqnarray}
The existence of conserved currents was naturally expected from the
fact that the left-right symmetric form of regularization preserves
global chiral symmetry, as pointed out in [22]. However, the trouble
observed in [22] was that these currents cannot be identified with
the currents probed by the external electroweak gauge fields.
This is related to the fact that the effective lagrangian in [22] with
the left-right symmetric regularization scheme,
breaks electromagnetic gauge invariance. In our present effective
lagrangian, this trouble has now been remedied owing to the
function of the newly subtracted local counter term.
In fact, the hadronic electroweak
currents are defined by
\begin{eqnarray}
   J^L_\mu \ \ \ \equiv \ \ \ \frac{\delta}{\delta l^\mu} \,\,
   {\Gamma_{eff} \,\,\,\vert}_{\,l_\mu,\,r_\mu \longrightarrow 0} \,\, ,\\
   J^R_\mu \ \ \ \equiv \ \ \ \frac{\delta}{\delta r^\mu} \,\,
   {\Gamma_{eff} \,\,\,\vert}_{\,l_\mu,\,r_\mu \longrightarrow 0} \,\, .
\end{eqnarray}
Here, $\Gamma_{eff}$ is our total effective action (in a special
gauge) given as
\begin{equation}
   \Gamma \ \ \ = \ \ \ \Gamma^{(n)} \ \ + \ \ \Gamma^{(a)} \,\, ,
\end{equation}
where
\begin{eqnarray}
   \Gamma^{(n)} \ \ &=& \ \ - \,i \,\,N_c \,\,\log \,\,\mbox{det} \,\,
   | \,D(\,U \,; \,{\cal L} , \,{\cal R} \,) \,| \nonumber \\
   &\,& \ - \ 
   \int \,\,\,d^4 x \,\,\,\frac{M^2_V}{2 \,g^2_V} \,\,\mbox{tr} \,\,
   [\,{({\cal L}_\mu \, - \,l_\mu)}^2 \ + \ 
   {({\cal R}_\mu \, - \,r_\mu)}^2
   \,] \,\, \ \ \\
   \Gamma^{(a)} \ \ &=& \ \ \Gamma^{(WZ)}_{LR} \,
   [\,U \, ; \,{\cal L},\,{\cal R} \,] \ - \ 
   \Gamma^{(a)}_{c.t.} \,[\,{\cal L},\,{\cal R} \,; \,l, \,r \,] \ - \ 
   \Gamma^{(WZ)}_{LR} \,[\,U = 1 \,; \,l, \,r \,] \,\, ,
\end{eqnarray}
with
\begin{eqnarray}
   \Gamma^{(a)}_{c.t.} \,\,[\,{\cal L},\,{\cal R} \,; \,l,\,r \,]
   \ \ \,\,\,&=& \ \ - \,\,\frac{c'}{2}
   \,\,\int_{S^4} \,\,\mbox{tr} \,\,
   [ \,\,{\cal L} \,\,l^3 \ + \ {\cal L} \,\,\{ \,dl, l \,\}
   \ + \ \frac{1}{2} \,\,{\cal L} \,\,l \,\,{\cal L} \,\,l
   \nonumber \\
   &\,& \hspace{5mm} + \,\,
   \{ \,d \,{\cal L}, \,{\cal L} \,\} \,\,l
   \ + \ {{\cal L}}^3 \,\,l \,\,] \ \ - \ \ 
   ( \,{\cal L} \leftrightarrow {\cal R}, \,l \leftrightarrow r \,)
   \,\,\, ,\\ \ \ \ \ 
   \Gamma^{(WZ)}_{LR} \,\,[\,U=1 \,; \,l,\,r \,]
   \ \ &=& \ \ - \,\,\frac{c'}{2}
   \,\,\int_{S^4} \,\,\mbox{tr} \,\,
   [ \,\,(\,l \,r \,- \,r \,l \,) \,\,(\,{\cal F}_l \, + \, {\cal F}_r \,)
   \nonumber \\
   &\,& \hspace{30mm} - \,l^3 \,r \ + \ r^3 \,l \ + \ 
   \frac{1}{2} \,l \,r \,l \,r \,\,] \,\,\, ,
\end{eqnarray}
with ${\cal F}_l \,= \,d \,l \,+ \,l^2$, ${\cal F}_r \,= \,d \,r \,+ \,r^2$.
Performing the functional derivative on $l^\mu$ and $r^\mu$ and then
letting $l_\mu$ and $r_\mu$ be zero, we find that
\begin{eqnarray}
   J^L_\mu \ \ &\equiv& \ \ \frac{M^2_V}{g^2_V} \,\,\,{\cal L}_\mu
   \ \ + \ \ \Delta_\mu \,\,({\cal L}_\mu) \,\, ,\\
   J^R_\mu \ \ &\equiv& \ \ \frac{M^2_V}{g^2_V} \,\,{\cal R}_\mu
   \ \ - \ \ \Delta_\mu \,\,({\cal R}_\mu) \,\, .
\end{eqnarray}
which precisely coincide with ${\bar{J}}^L_\mu$ and ${\bar{J}}^R_\mu$,
the conservation of which we have already proved.
We are thus led to complete CVC and CAC relations as follows :
\begin{equation}
   \partial^\mu \,J^V_\mu \ \ = \ \ 0 \, ,\ \ \ \ 
   \partial^\mu \,J^A_\mu \ \ = \ \ 0 \, ,
\end{equation}
with the definition $J^V_\mu = \frac{1}{2} \,(J^R_\mu + J^L_\mu)$ and
$J^A_\mu = \frac{1}{2} \,(J^R_\mu - J^L_\mu)$.
Eqs.(3.75) and (3.76) shows, as first pointed out
by Bijnens and Prades [32],
that the exact current-field identity is lost in the new scheme.
Note however that it is only minimally modified. Since the deviation
from the current-field identity depends on the vector and axial-vector
fields only, the external electroweak gauge fields
are coupled to the Goldstone bosons only through the hadronic vector
and axial-vector fields. The vector (and axial-vector) meson dominance
still holds in this narrow sense.

\vspace{0mm}
\setcounter{equation}{0}
\section{Summary and Discussion}

\ \ \ \ \ \ Using the standard auxiliary field method, we have derived
from the extended NJL model an effective meson action, which contains
not only the Nambu-Goldstone bosons but also the vector and axial-vector
mesons. The obtained effective action consists of the nonanomalous
(intrinsic parity conserving) part and the anomalous (intrinsic parity
violating ) part. The nonanomalous part just coincides with the
lagrangian of Bando et al. obtained on the basis of the enlarged hidden
local symmetry, except that some of the parameters in their model
lagrangian cannot be arbitrary in our effective lagrangian derived
from the extended NJL model. A notable feature of our effective
action is that not only the nonanomalous part
but also the anomalous part is completely
invariant under the enlarged hidden local transformation
$(h_L (x), h_R (x)) \in {[U(n)_L \times U(n)_R]}^{(HLS)}$.
Putting it in another way, there is no gauge anomaly in the enlarged hidden
local symmetry. From the physical viewpoint, however, the most important
symmetry of an effective action of QCD is the global chiral symmetry.
If we switch off the couplings with the external gauge fields, the
anomalous action that satisfies this property can easily be obtained by
choosing the left-right symmetric form of regularization scheme.
However, once the electroweak couplings are introduced, there arises a
nontrivial problem. This is because naive use of the left-right symmetric
form of regularization breaks the electromagnetic gauge invariance.
To maintain the global chiral symmetry of the strong interaction together
with the electromagnetic gauge invariance, we need to subtract counter
terms, which depend on both the hadronic vector and axial-vector fields
and the external gauge fields. This renormalization procedure
enables us to obtain an effective action, which respects the global chiral
symmetry at the strong interaction level as well as the electromagnetic
gauge invariance, while keeping the full hidden local symmetry
${[U(n)_L \times U(n)_R]}^{(HLS)}$. 

In this process of constructing an effective action
consistent with the symmetries of QCD, it has become clear that the
concept of the hidden local symmetry plays no positive role, which makes us
to reconfirm several authors' suspicion that it may not
be a physical symmetry [4,43]. Remember that, in our derivation of the
action with enlarged hidden local symmetry, these extra gauge degrees of
freedom are introduced by hand with the inclusion of two kinds of
compensating fields (or ``compensators'').
There are several familiar examples of such compensating mechanism.
A classical example is the scalar of the Stueckelberg formalism, which
is used to introduce a local gauge invariance into a theory with a
massive vector fields [44]. The scalar field of the chiral Schwinger model
(in $1 + 1$ space-time dimension), which is introduced so as to cancel
the chiral anomaly of the original theory [45],
may also be thought of as a kind
of compensator. The role of the hidden local symmetry and the associated
compensating fields was discussed by de Wit and Grisaru in quite a
general context [46]. Their general argument goes as follows.
At the classical level, a theory with the extra gauge degrees of freedom is
completely equivalent to the original theory, since the compensators can
always be gauged away via the gauge transformation.
Interestingly, the same is true also for theories that can be consistently
quantized, since classically irrelevant gauge degrees of freedom also
decouple at the quantum level, as a consequence of the Ward identities
(or BRST invariance) corresponding to the classical gauge symmetry.
This means that theories described with and without compensators are
physically equivalent. There is one caveat in the above reasoning, however.
If anomalies are present in the gauge symmetries in question, the theory
becomes anomalous, i.e. it cannot be consistently quantized ; unitarity
is violated, the gauge degrees of freedom no longer decouple etc.
It is clear from the discussion so far that our effective action has
no such inconsistency.
Its anomalous part has been obtained as a straightforward
natural of the gauged Wess-Zumino-Witten action. Owing to the
function of the extra dynamical fields, i.e. the compensators, the
potentially dangerous anomaly never appears.
According to the expression by
de Wit and Grisaru, the compensators also compensate anomaly !

At any rate, since our final effective action is completely hidden gauge
invariant, the extra gauge degrees of freedom carried by the compensators
can always be gauged away and decouple from all physical processes.
Why ever do we consider such unphysical symmetries, then ?
There are several advantages in working in a theory with extra gauge
degrees of freedom. By moving freely from one gauge to another,
one can get a unified view of the seemingly independent ideas.
For example, we have seen that the massive Yang-Mills scheme with
the (approximate) VMD
type couplings with the external electroweak gauge fields can be
regarded as a gauge fixed version of a lagrangian with enlarged hidden
local symmetry at least formally, while we can simultaneously arrive
at a clear understanding that the idea of massive Yang-Mills scheme, i.e.
the {\it full gauging} of the global chiral symmetry has no theoretical
foundation [47].

We also recall the fact that by introducing the extra gauge symmetries the
chiral symmetries are linearly realized. 
Usefulness of this property may, for example, be deduced from the
observation that the standard description of photon by means of a vector
potential, rather than two transverse degrees of freedom, may be viewed
as resulting from the introduction of a compensating mechanism used to
linearlize its Lorentz transformation [46].
It is also expected to play useful roles when
quantizing the theory to evaluate meson loop diagrams.
Unfortunately, our effective action, though
it can be consistently quantized, is not renormalizable in the usual
power-counting sense. Recently, Gomis and Weinberg argue [48] that
some gauge theories, that are not renormalizable in Dyson's sense,
may nevertheless be
renormalizable in the modern sense that all the divergences can be
eliminated by renormalization of the infinite number of terms in the bare
action. It is an interesting open question whether the concept of hidden
local symmetry in effective theories of QCD may play some useful role in
the context of this generalized interpretation of renormalizable theories.

\vspace{6mm}
\appendix
\section{Appendix}%

\ \ \ \ \ \ The effective action $\Gamma^{(a)}_{LR}$ derived in sect.2
can trivially be generalized to arbitrary even-dimensional space-time
case ($n = D / 2$ with $D$ being the space-time dimension) as
\begin{eqnarray}
   \Gamma^{(a)}_{LR} \ 
   &=& \ c_n \,\,\int_{B^{2n+1}} \,\,\,[ \,\,
   \omega^0_{2n+1} \,( 0, \,\xi^\dagger_R \,d \xi_R) \ - \ 
   \omega^0_{2n+1} \,( 0, \,\xi^\dagger_L \,d \xi_L) \ - \ 
   \omega^0_{2n+1} \,( 0, \,\xi_M \,d \,\xi^\dagger_M) \,\,] \ \nonumber \\
   &+& \ c_n \,\,\int_{S^{2n}} \,\,\,\,\,[ \,\,
   \rho_{2n} \,(0, \,R, \,\xi^\dagger_R \,d \,\xi_R) \ + \ 
   \rho_{2n} \,(0, \,\xi^\dagger_L \,d \,\xi_L, \,L) \nonumber \\
   &\,& \hspace{22mm} + \ \ 
   \rho_{2n} \,(0, \,\xi_M \,d \,\xi^\dagger_M, \,\tilde{L}) \ - \ 
   \rho_{2n} \,(0, \,\hat{L}, \,\hat{R}) \,\,] \,\, ,
\end{eqnarray}
with $c_n = {(- \,i)}^{n+1} \,N_c \,/ \,{(2 \pi)}^n \,(n+1)!$.
The responses of the above action under an arbitrary gauge variations
can easily be evaluated by using the method described in [39].
First, we show the response of each term of (A.1) under the
${[U(n)_L \times U(n)_R]}^{(HLS)}$ transformation. Under the infinitesimal
left variation ($h_L = e^{- \,\epsilon_L} \simeq 1 - \epsilon_L,
h_R = 1$), we find that
\begin{eqnarray}
   \delta^{(HLS)}_L \,\,\omega^0_{2n+1} \,(0,\,\xi^\dagger_R \,d \,\xi_R)
   \ &=& \ \ 0 \,\, , \\
   \delta^{(HLS)}_L \,\,\omega^0_{2n+1} \,(0,\,\xi^\dagger_L \,d \,\xi_L)
   \ &=& \ {(-1)}^{n+1} \,\,\frac{(n+1) \,{(n!)}^2}{(2 \,n)!} \,\,
   d \,\,\mbox{tr} \,\,\epsilon_L \,\,
   {(\xi_L \,d \,\xi^\dagger_L)}^{2n} \,\, ,\\
   \delta^{(HLS)}_L \,\,\omega^0_{2n+1} \,(0,\,\xi_M \,d \,\xi^\dagger_M)
   \ &=& \ {(-1)}^{n} \,\,\frac{(n+1) \,{(n!)}^2}{(2 \,n)!} \,\,
   d \,\,\mbox{tr} \,\,\epsilon_L \,\,
   {(\xi_M \,d \,\xi^\dagger_M)}^{2n} \,\, ,\\
   \delta^{(HLS)}_L \,\,\rho_{2n} \,(0,\,R,\,\xi^\dagger_R \,d \,\xi_R)
   \ &=& \ \ 0 \,\, , \\
   \delta^{(HLS)} \,\,\rho_{2n} \,(0,\,\xi^\dagger_L \,d \,\xi_L,\,L)
   \ &=& \ {(-1)}^{n} \,\,\frac{(n+1) \,{(n!)}^2}{(2 \,n)!} \,\,
   \mbox{tr} \,\,\epsilon_L \,\,
   {(\xi_L \,d \,\xi^\dagger_L)}^{2n}
   \ - \ \bar{\Delta} (\epsilon_L,\tilde{L}) \,\, , \ \ \ \ \\
   \delta^{(HLS)}_L \,\,\rho_{2n} \,(0,\,\xi_M \,d \,\xi^\dagger_M,
   \,\tilde{L})
   \ &=& \ {(-1)}^{n} \,\,\frac{(n+1) \,{(n!)}^2}{(2 \,n)!} \,\,
   \mbox{tr} \,\,\epsilon_L \,\,
   {(\xi_M \,d \,\xi^\dagger_M)}^{2n}
   \ + \ \bar{\Delta} (\epsilon_L,\tilde{L}) \,\, , \ \ \ \ \ \\
   \delta^{(HLS)}_L \,\,\rho_{2n} \,(0,\,\hat{L},\,\hat{R})
   \ &=& \ \ 0 \,\, .
\end{eqnarray}
Here we have defined the quantity :
\begin{equation}
   \bar{\Delta} \,(\theta, A) \ = \ (n+1) \,\,\sum_{p=0}^{n-1} \,\,
   \int_0^1 \,\,ds \,\,(1 - s) \,\,\mbox{tr} \,\,
   [ \,d \,\theta \,\,{( s \,dA + s^2 A^2 )}^p \,\,A \,\,
   {( s \,dA + s^2 A^2 )}^{n-p-1} \,] \,\, .
\end{equation}
For $n = D / 2 = 2$, this reduces to
\begin{equation}
   \bar{\Delta} \,(\theta, A) \ \,= \ \ \mbox{tr} \,\,
   d \,\theta \,\,\,\frac{1}{2} \,\,
   ( \,A \,d A \, + \, d A \,A \, + \,A^3 \,) \,\, .
\end{equation}
On the other hand, under the infinitesimal right variation ($h_L = 1,
h_R = e^{- \,\epsilon_R} \simeq 1 - \epsilon_R$), we have
\begin{eqnarray}
   \delta^{(HLS)}_R \,\,\omega^0_{2n+1} \,(0,\,\xi^\dagger_R \,d \,\xi_R)
   \ &=& \ {(-1)}^{n+1} \,\,\frac{(n+1) \,{(n!)}^2}{(2 \,n)!} \,\,
   d \,\,\mbox{tr} \,\,\epsilon_R \,\,
   {(\xi_R \,d \,\xi^\dagger_R)}^{2n} \,\, ,\\
   \delta^{(HLS)}_R \,\,\omega^0_{2n+1} \,(0,\,\xi^\dagger_L \,d \,\xi_L)
   \ &=& \ \ 0 \,\, , \\
   \delta^{(HLS)}_R \,\,\omega^0_{2n+1} \,(0,\,\xi_M \,d \,\xi^\dagger_M)
   \ &=& \ {(-1)}^{n+1} \,\,\frac{(n+1) \,{(n!)}^2}{(2 \,n)!} \,\,
   d \,\,\mbox{tr} \,\,\epsilon_R \,\,
   {(\xi^\dagger_M \,d \,\xi_M)}^{2n} \,\, ,\\
   \delta^{(HLS)}_R \,\,\rho_{2n} \,(0,\,R,\,\xi^\dagger_R \,d \,\xi_R)
   \ &=& \ {(-1)}^{n} \,\,\frac{(n+1) \,{(n!)}^2}{(2 \,n)!} \,\,
   \mbox{tr} \,\,\epsilon_R \,\,
   {(\xi_R \,d \,\xi^\dagger_R)}^{2n}
   \ + \ \bar{\Delta} (\epsilon_R,\tilde{R}) \,\, ,\ \ \ \ \ \\
   \delta^{(HLS)}_R \,\,\rho_{2n} \,(0,\,\xi^\dagger_L \,d \,\xi_L,\,L)
   \ &=& \ \ 0 \,\, , \\
   \delta^{(HLS)}_R \,\,\rho_{2n} \,(0,\,\xi_M \,d \,\xi^\dagger_M,
   \,\tilde{L})
   \ &=& \ {(-1)}^{n+1} \,\,\frac{(n+1) \,{(n!)}^2}{(2 \,n)!} \,\,
   \mbox{tr} \,\,\epsilon_R \,\,
   {(\xi^\dagger_M \,d \,\xi_M)}^{2n}
   \, - \ \bar{\Delta} (\epsilon_R,\hat{L}) \,\, ,\ \ \ \ \ \ \\
   \delta^{(HLS)}_R \,\,\rho_{2n} \,(0,\,\hat{L},\,\hat{R})
   \ &=& \ \bar{\Delta} (\epsilon_R,\tilde{R}) \ - \ 
   \bar{\Delta} (\epsilon_R, \hat{L})  \,\, .
\end{eqnarray}
One can easily convince that, for either of the right or left hidden
gauge variation, the responses of the individual terms cancel out to
be zero, thereby leading to the result :
\begin{equation}
   \delta^{(HLS)}_L \,\,\Gamma^{(a)}_{LR} \ \ = \ \ 
   \delta^{(HLS)}_R \,\,\Gamma^{(a)}_{LR} \ \ = \ \ 0 \,\, ,
\end{equation}
which denotes that $\Gamma^{(a)}$ is completely invariant under the
enlarged hidden local symmetry.

We can similarly evaluate the response of each term of $\Gamma^{(a)}$
under the external gauge variations belonging to
${[U(n)_L \times U(n)_R]}^{(ext)}$. Under the infinitesimal left variation
($g_L = e^{- \,\theta_L} \simeq 1 - \theta_L, g_R = 1$), we find that
\begin{eqnarray}
   \delta^{(ext)}_L \,\,\omega^0_{2n+1} \,(0,\,\xi^\dagger_R \,d \,\xi_R)
   \ &=& \ \ 0 \,\, , \\
   \delta^{(ext)}_L \,\,\omega^0_{2n+1} \,(0,\,\xi^\dagger_L \,d \,\xi_L)
   \ &=& \ {(-1)}^{n} \,\,\frac{(n+1) \,{(n!)}^2}{(2 \,n)!} \,\,
   d \,\,\mbox{tr} \,\,\theta_L \,\,
   {(\xi^\dagger_L \,d \,\xi_L)}^{2n} \,\, ,\\
   \delta^{(ext)}_L \,\,\omega^0_{2n+1} \,(0,\,\xi_M \,d \,\xi^\dagger_M)
   \ &=& \ \ 0 \,\, ,\\
   \delta^{(ext)}_L \,\,\rho_{2n} \,(0,\,R,\,\xi^\dagger_R \,d \,\xi_R)
   \ &=& \ \ 0 \,\, ,\ \\
   \delta^{(ext)} \,\,\rho_{2n} \,(0,\,\xi^\dagger_L \,d \,\xi_L,\,L)
   \ &=& \ {(-1)}^{n} \,\,\frac{(n+1) \,{(n!)}^2}{(2 \,n)!} \,\,
   \mbox{tr} \,\,\theta_L \,\,
   {(\xi^\dagger_L \,d \,\xi_L)}^{2n}
   \ + \ \bar{\Delta} (\theta_L,L) \,\, ,\ \ \ \ \ \\
   \delta^{(ext)}_L \,\,\rho_{2n} \,(0,\,\xi_M \,d \,\xi^\dagger_M,
   \,\tilde{L})
   \ &=& \ \ 0 \,\, ,\\
   \delta^{(ext)}_L \,\,\rho_{2n} \,(0,\,\hat{L},\,\hat{R})
   \ &=& \ \ 0 \,\, .
\end{eqnarray}
On the other hand, the infinitesimal right variation ($g_L = 1, g_R =
e^{- \,\theta_R} \simeq 1 - \theta_R$) gives
\begin{eqnarray}
   \delta^{(ext)}_R \,\,\omega^0_{2n+1} \,(0,\,\xi^\dagger_R \,d \,\xi_R)
   \ &=& \ {(-1)}^{n} \,\,\frac{(n+1) \,{(n!)}^2}{(2 \,n)!} \,\,
   d \,\,\mbox{tr} \,\,\theta_R \,\,
   {(\xi^\dagger_R \,d \,\xi_R)}^{2n} \,\, ,\\
   \delta^{(ext)}_R \,\,\omega^0_{2n+1} \,(0,\,\xi^\dagger_L \,d \,\xi_L)
   \ &=& \ \ 0 \,\, ,\\
   \delta^{(ext)}_R \,\,\omega^0_{2n+1} \,(0,\,\xi_M \,d \,\xi^\dagger_M)
   \ &=& \ \ 0 \,\, ,\\
   \delta^{(ext)}_R \,\,\rho_{2n} \,(0,\,R,\,\xi^\dagger_R \,d \,\xi_R)
   \ &=& \ {(-1)}^{n+1} \,\,\frac{(n+1) \,{(n!)}^2}{(2 \,n)!} \,\,
   \mbox{tr} \,\,\theta_R \,\,
   {(\xi^\dagger_R \,d \,\xi_R)}^{2n}
   \ - \ \bar{\Delta} (\theta_R,R) \,\, ,\ \ \ \ \ \\
   \delta^{(ext)}_R \,\,\rho_{2n} \,(0,\,\xi^\dagger_L \,d \,\xi_L,\,L)
   \ &=& \ \ 0 \,\, ,\\
   \delta^{(ext)}_R \,\,\rho_{2n} \,(0,\,\xi_M \,d \,\xi^\dagger_M,
   \,\tilde{L})
   \ &=& \ \ 0 \,\, ,\\
   \delta^{(ext)}_R \,\,\rho_{2n} \,(0,\,\hat{L},\,\hat{R})
   \ &=& \ \ 0 \,\, .
\end{eqnarray}
Summing up all the contributions, we are led to the expected result,
i.e. the left-right symmetric form of anomaly given as,
\begin{eqnarray}
   \delta^{(ext)}_L \,\,\Gamma^{(a)}_{LR} \ \ &=& \ \ \,\,\,\,\,\,\,\,
   c_n \,\,\,\int_{S^4} \,\,\,\bar{\Delta} \,(\theta_L, L) \,\, , \\
   \delta^{(ext)}_R \,\,\Gamma^{(a)}_{LR} \ \ &=& \ \ - \,\,\,
   c_n \,\,\,\int_{S^4} \,\,\,\bar{\Delta} \,(\theta_R, R) \,\, .
\end{eqnarray}
Note that $\bar{\Delta} \,(\theta, A)$ is proportional to $d \, \theta$
(see (A.9) or (A.10)).
It is then obvious that, specializing to the global chiral
transformation, the right hand sides of (A.33) and (A.34) both vanish.

\newpage
%

\vspace{8mm}
\begin{flushleft}
\large\bf{Figure caption} \\
\end{flushleft}
\ \\
\begin{minipage}{2cm}
Fig. 1(a).
\end{minipage}
\begin{minipage}[t]{13cm}
The integral path in the field space for obtaining the anomalous action
with the symmetry $G_{global} \times G_{local}$ with
$G_{global} = {[U(n)_L \times U(n)_R]}^{(ext)}$ and
$G_{local} = {[U(n)_L \times U(n)_R]}^{(HLS)}$.
\end{minipage}
\ \\
\vspace{6mm}
\ \\
\begin{minipage}{2cm}
Fig. 1(b).
\end{minipage}
\begin{minipage}[t]{13cm}
The decomposition of the integral path of fig.1(a) into two parts, which
respectively correspond to the contributions from the Jacobians $J_1$
and $J_2$ in (2.65).
\end{minipage}

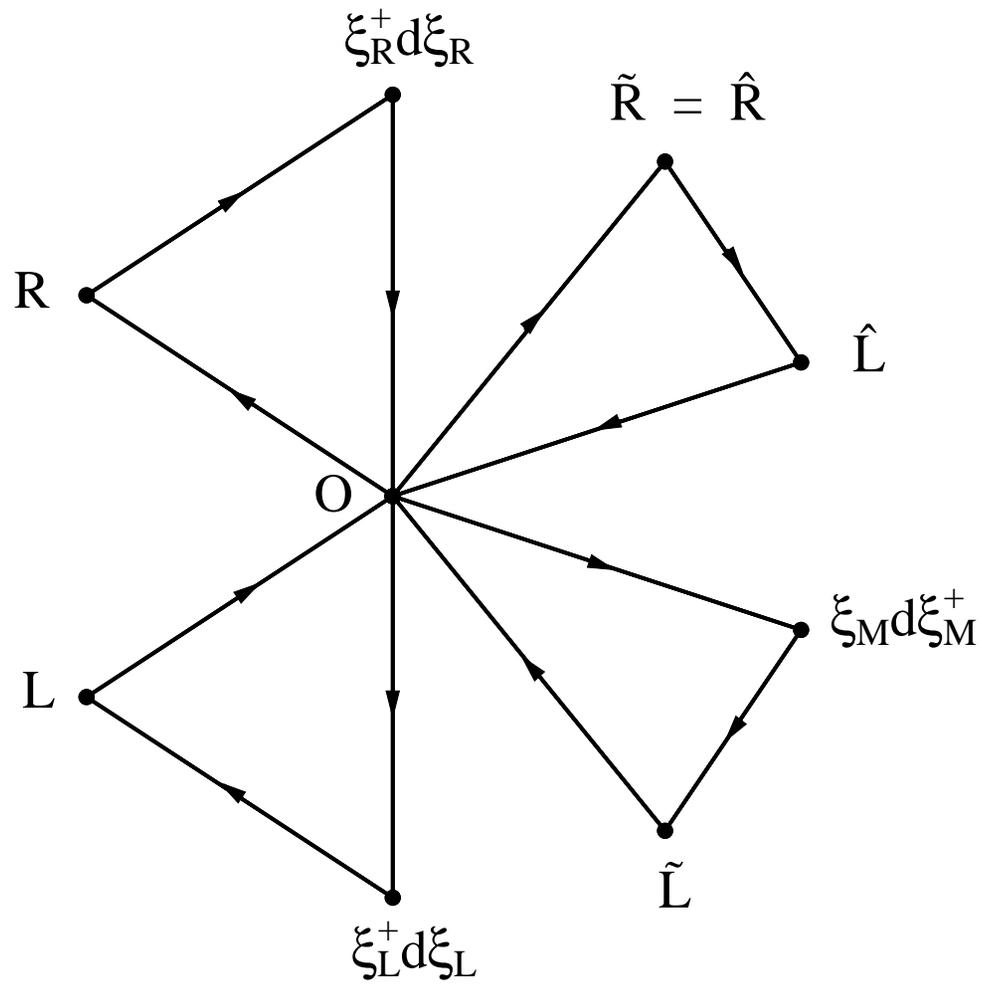
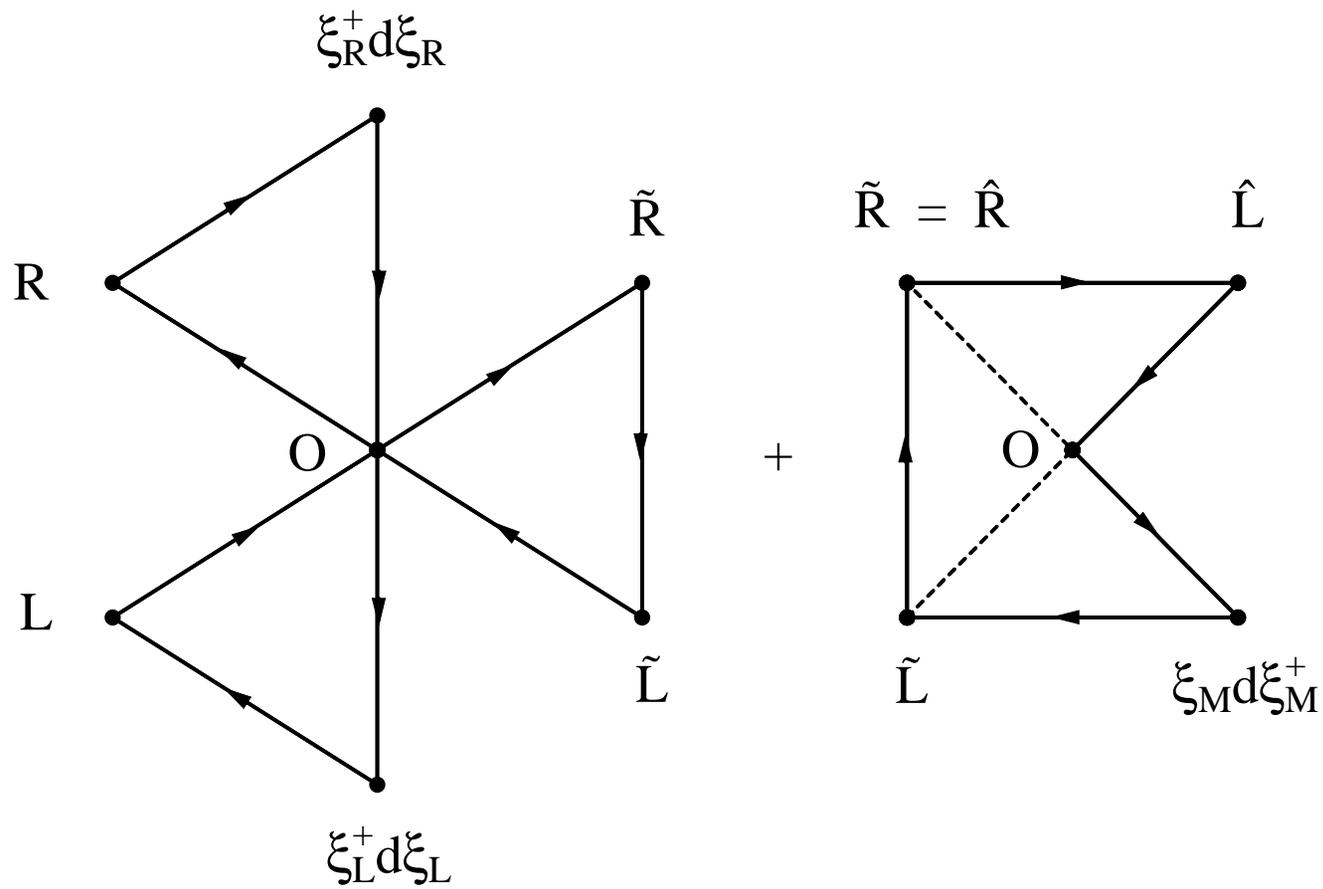
\end{document}